\documentclass[12pt]{article}
\usepackage{epsf}
\usepackage{cite}
\newcommand{\sss}{\scriptscriptstyle}
\begin{document}
\begin{titlepage}
\begin{flushright}FSU-HEP-2001-0602
\end{flushright} 
\begin{flushright}BNL-HET-01/19
\end{flushright}
\begin{flushright}UR-1639
\end{flushright}
\begin{flushright}hep-ph/0109066
\end{flushright}
\vspace{1.truecm}
\begin{center}
{\large\bf
QCD Corrections to Associated\\
\vskip .2in 
$t\bar t h$ Production at the Tevatron}
\\
\vspace{1in}
{\bf L.~Reina}\\
{\it  Physics Department, Florida State University,\\ 
Tallahassee, FL 32306-4350, USA}
\\ 
\vspace{.25in}
{\bf S.~Dawson}\\
{\it Physics Department, Brookhaven National Laboratory,\\
Upton, NY 11973, USA}
\\ 
\vspace{.25in}
{\bf D.~Wackeroth}\\
{\it Department of Physics and Astronomy, University of Rochester,\\
Rochester, NY 14627-0171, USA}
\vspace{1in}  
\end{center}
\begin{abstract} 
  We present in detail the calculation of the ${\cal O}(\alpha_s^3)$
  inclusive total cross section for the process $p {\overline p}
  \rightarrow t {\overline t} h$ in the Standard Model, at the
  Tevatron center-of-mass energy $\sqrt{s_{\sss H}}\!=\!2$~TeV.  The
  next-to-leading order QCD corrections significantly reduce the
  renormalization and factorization scale dependence of the Born cross
  section. They slightly decrease or increase the Born cross section
  depending on the values of the renormalization and factorization
  scales.\clearpage
\end{abstract}
\end{titlepage} 
\section{Introduction}
\label{sec:intro}
Among the most important goals of present and future colliders is the
study of the electroweak symmetry breaking mechanism and the origin of
fermion masses.  If the introduction of one or more Higgs fields is
responsible for the breaking of the electroweak symmetry, then at
least one Higgs boson should be relatively light, and certainly in the
range of energies of present (Tevatron) or future (LHC) hadron
colliders.  The present lower bounds on the Higgs boson mass from
direct searches at LEP2 are $M_h\!>\!114.1$~GeV (at $95\%$ CL) 
\cite{lephwg1} for the Standard Model (SM) Higgs boson, and
$M_{h^0}\!>\!91.0$~GeV and $M_{A^0}\!>\!91.9$~GeV (at $95\%$ CL,
$0.5\!<\!\tan\beta\!<\!2.4$ excluded) \cite{lephwg2} for the light
scalar ($h^0$) and pseudoscalar ($A^0$) Higgs bosons of the minimal
supersymmetric standard model (MSSM).  At the same time, global SM
fits to all available electroweak precision data indirectly point to
the existence of a light Higgs boson, $M_h\!<\!212-236$~GeV
\cite{lepewwg}, while the MSSM requires the existence of a scalar
Higgs boson lighter than about 130~GeV
\cite{Heinemeyer:1999np}. Therefore, the possibility of a Higgs boson
discovery in the mass range around 115-130~GeV seems increasingly
likely.

In this context, the Tevatron will play a crucial role and can
potentially discover a Higgs boson in the mass range between the
present experimental lower bound and about 180~GeV
\cite{Carena:2000yx}. The dominant Higgs production modes at the
Tevatron are gluon-gluon fusion ($gg\rightarrow h$) and the associated
production with a weak boson ($q\bar q\rightarrow Wh,~Zh$).  Because
of small event rates and large backgrounds, the Higgs boson search in these
channels is extremely difficult, requiring the highest possible
luminosity.  It is therefore important to investigate all possible
production channels, in the effort to fully exploit the range of
opportunities offered by the available statistics.

Recently, attention has been drawn to the possibility of detecting a
Higgs signal in association with a pair of top-antitop quarks at the
Tevatron, i.e. in $p\bar p\rightarrow t\bar th$
\cite{Goldstein:2001bp}. This production mode can play a role over
most of the Higgs mass range accessible at the Tevatron. Although it
has a small event rate, $\sim 1-5$~fb for a SM-like Higgs boson, and even
lower for an MSSM Higgs boson, the signature ($W^+W^- b \bar b b\bar
b$) is quite spectacular.  Furthermore, at the Tevatron, after fully
reconstructing both top quarks, the shape of the invariant mass
distribution of the remaining $b {\overline b}$ pair is quite
different for the signal and for the background.  The statistics is
too low to allow any direct measurement of the top-quark Yukawa
coupling, but recent studies \cite{incandela} indicate that this
channel can reduce the luminosity required for the discovery of a
SM-like Higgs boson at Run II of the Tevatron by as much as 15-20\%.

The total cross section for $p\bar p\rightarrow t\bar th$ has been
known at tree-level for quite some time \cite{Kunszt:1984ri}.  As for
any other hadronic process, next-to-leading (NLO) QCD corrections are
expected to be important and are crucial in order to reduce the
dependence of the cross section on the renormalization and
factorization scales.  Preliminary indications of the magnitude of the
NLO QCD corrections can be obtained in the framework of the Effective
Higgs Approximation (EHA), where terms of order $M_h/\sqrt{s}$ and
$M_h/m_t$ are systematically neglected in the computation
\cite{Dawson:1998im}. This approximation correctly reproduces the
collinear bremsstrahlung of the Higgs boson from the heavy top quarks.
However, we expect the EHA to be more reliable at the LHC
center-of-mass energies, for which it was originally proposed, than at
the Tevatron center-of-mass energies. We will briefly discuss the
predictions of the EHA for $p\bar p\rightarrow t\bar t h$ in
Sec.~\ref{sec:results}.

We also notice that QCD corrections to the associated production of a
Higgs boson with a pair of $b\bar b$ quarks, which is dominated by the
$gg\rightarrow b\bar b h$ channel, have been computed in the limit of
large $M_h$ \cite{Dicus:1989cx}, by resuming the leading
$\ln(M_h/m_b)$ terms. However this result cannot be applied to the $t
{\overline t} h$ production of a relatively light Higgs boson at the
Tevatron, both because the ratio $M_h/m_t$ is of ${\cal O}(1)$ and
does not justify the large $M_h$ limit, and because the $gg$ channel
is negligible for $t\bar th$ production at the Tevatron.

\begin{figure}[bt]
\centering
\epsfxsize=5.in
\leavevmode\epsffile{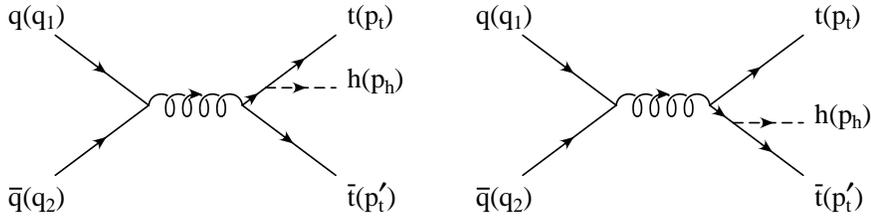}
\caption[]{Feynman diagrams contributing to the lowest
  order process, $q\bar q\rightarrow t\bar th$. The arrows indicate
  the momentum flow.}
\label{fig:qqtth_tree}
\end{figure}

In this paper we present in detail the calculation of the NLO 
inclusive total cross section for $p\bar p\rightarrow t\bar th$,
$\sigma_{\sss NLO}(p\bar p\rightarrow t\bar th)$, in the Standard
Model, at the Tevatron center-of-mass energy. For $p {\overline p}$
collisions at hadronic center-of-mass energy $\sqrt{s_{\sss
    H}}\!=\!2$~TeV, more than $95\%$ of the tree-level cross section
comes from the sub-process $q \bar q\rightarrow t\bar th$.  Therefore,
we include only the $q\bar q\rightarrow t\bar th$ channel when
computing the tree-level total cross section, and we calculate the NLO
total cross section by adding the complete set of virtual and real
${\cal O}(\alpha_s)$ corrections to $q\bar q\rightarrow t\bar th$. The
Feynman diagrams contributing to $q\bar q\rightarrow t\bar th$ at
lowest order are shown in Fig.~\ref{fig:qqtth_tree}, while examples of
${\cal O}(\alpha_s)$ virtual and real corrections are given in
Figs.~\ref{fig:selfenergies}-\ref{fig:real}.  The main challenge in
the calculation of the ${\cal O}(\alpha_s)$ virtual corrections comes
from the presence of pentagon diagrams with several massive external
and internal particles. We have calculated the corresponding pentagon
scalar integrals as linear combinations of scalar box integrals using
the method of Ref.~\cite{Bern:1993em}.  The real corrections are
computed using the phase space slicing method, in both the double
\cite{bergmann,Harris:2001sx} and single
\cite{Giele:1992vf,Giele:1993dj,Keller:1999tf} cutoff approach.  This
is the first application of the single cutoff phase space slicing
approach to a cross section involving more than one massive particle
in the final state.

Numerical results for our calculation of $\sigma_{\sss NLO}(p\bar
p\rightarrow t\bar th)$ at the hadronic center-of-mass energy
$\sqrt{s_{\sss H}}\!=\!2$~TeV have been presented in
\cite{Reina:2001sf}.  An independent calculation of the NLO total
cross section for $p \bar p\rightarrow t\bar th$ has been performed by
Beenakker {\it et al.}~\cite{Beenakker:2001rj}.  The numerical results
of both calculations have been compared and they are found to be in
very good agreement.  The ${\cal O}(\alpha_s)$ corrections to the
sub-process $g g \rightarrow t {\overline t} h$ are, however, crucial
for determining $\sigma_{\sss NLO}(pp\rightarrow t\bar th)$ at the
LHC, since $pp$ collisions at $\sqrt{s_{\sss H}}\!=\!14$ TeV are
dominated by the gluon-gluon initial state.  Results for the LHC are
presented elsewhere \cite{Beenakker:2001rj,ggtth}.

The outline of our paper is as follows. In Sec.~\ref{sec:framework} we
summarize the general structure of the NLO cross section, and proceed
in Secs.~\ref{sec:virtual} and \ref{sec:real} to present the details
of the calculation of both the virtual and real parts of the NLO QCD
corrections. In Sec.~\ref{sec:total} we explicitly show the
factorization of the initial-state singularities into the quark
distribution functions, and finally summarize our result for the NLO
inclusive total cross section for $p\bar p \rightarrow t \bar th$ at
the Tevatron in Eqs.~(\ref{eq:sigmatot2}) and (\ref{eq:sigmatot1}).
Numerical results for the total cross section are presented in
Sec.~\ref{sec:results}.  Explicit analytic results for the scalar
pentagon and the infrared-singular box integrals are presented in
Appendices~\ref{sec:scalar_pentagon} and \ref{sec:scalar_boxes}.
Appendix~\ref{sec:soft_int} contains a collection of soft phase space
integrals that are used in the calculation of the real ${\cal
  O}(\alpha_s)$ corrections to $q\bar q\rightarrow t\bar th$ with the
double cutoff phase space slicing method. Finally, in
Appendix~\ref{sec:a_real_a1a2a3a4} we give the explicit structure of
the real gluon emission color ordered amplitudes that are used in the
calculation of the real ${\cal O}(\alpha_s)$ corrections to $q\bar
q\rightarrow t\bar th$ with the single cutoff phase space slicing
method.

\section{General Framework}
\label{sec:framework}

The inclusive total cross section for $p\bar p\rightarrow t\bar th$ at
${\cal O}(\alpha_s^3)$ can be written as:
\begin{equation}
\label{eq:sigma_nlo}
\sigma_{\sss NLO}(p\bar p\rightarrow t\bar t h) 
=\sum_{ij}\int dx_1 dx_2 {\cal F}_i^p(x_1,\mu) {\cal F}_j^{\bar p}(x_2,\mu)
{\hat \sigma}^{ij}_{\sss NLO}(x_1,x_2,\mu)\,\,\,,
\end{equation}
where $ {\cal F}_i^{p, {\overline p}}$ are the NLO parton distribution
functions (PDF) for parton $i$ in a proton/antiproton, defined at a
generic factorization scale $\mu_f\!=\!\mu$, and ${\hat
  \sigma}^{ij}_{\sss NLO}$ is the ${\cal O}(\alpha_s^3)$ parton-level
total cross section for incoming partons $i$ and $j$, composed of the
two channels $q {\overline q}$, $gg\rightarrow t {\overline t}h$, and
renormalized at an arbitrary scale $\mu_r$ which we also take to be
$\mu_r\!=\!\mu$.  Throughout this paper we will always assume the
factorization and renormalization scales to be equal,
$\mu_r\!=\!\mu_f\!=\!\mu$.  The partonic center-of-mass energy
squared, $s$, is given in terms of the total hadronic center-of-mass
energy squared, $s_{\sss H}$, by $s=x_1 x_2 s_{\sss H}$.  As explained
in the introduction, we consider only the $q\bar q\rightarrow t\bar
th$ channel, summed over all light quark flavors, and neglect the
$gg\rightarrow t\bar t h$ channel, since the $gg$ initial state is
numerically irrelevant at the Tevatron.

We write the NLO parton-level total cross section ${\hat
  \sigma}_{\sss NLO}^{ij}( x_1,x_2,\mu)$ as:
\begin{eqnarray}
\label{eq:sigmahat_nlo}
{\hat \sigma}_{\sss NLO}^{ij}(x_1,x_2,\mu)&=&
\alpha_s^2(\mu)\biggl\{f_{\sss LO}^{ij}(x_1,x_2)+
{\alpha_s(\mu)\over 4 \pi}f_{\sss NLO}^{ij}(x_1,x_2,\mu)\biggr\}\\
&\equiv&{\hat \sigma}_{\sss LO}^{ij}(x_1,x_2,\mu)+
\delta {\hat \sigma}_{\sss NLO}^{ij}(x_1,x_2,\mu)\,\,\,,\nonumber
\end{eqnarray}
where $\alpha_s(\mu)$ is the strong coupling constant renormalized at
the arbitrary scale $\mu_r\!=\!\mu$, ${\hat \sigma}_{\sss LO}^{ij}(
x_1,x_2,\mu)$ is the ${\cal O}(\alpha_s^2)$ Born cross section, and
$\delta {\hat \sigma}_{\sss NLO}^{ij}( x_1,x_2,\mu)$ consists of the ${\cal
  O}(\alpha_s)$ corrections to the Born cross section, including
the effects of mass factorization (see Sec.~\ref{sec:total}). 

The Born cross section to $q {\overline q}
\rightarrow t {\overline t} h$  is given by \cite{Gaemers:1978jr}:
\begin{eqnarray}
{\hat \sigma}^{q\bar q}_{\sss LO}(x_1,x_2,\mu)&=& 
{\alpha_s^2(\mu)\over 27\pi s} 
\biggl({m_t\over v}\biggr)^2\int^{x_h^{max}}_{x_h^{min}} dx_h 
\biggl\{
{4 {\hat \beta}\over x_h^2-{\hat \beta}^2}
\biggl(1+{2m_t^2\over s}\biggr)\biggr({4 m_t^2-M_h^2\over s}
\biggr)+\\
\biggl[ x_h\!\!\!\!&+&\!\!\!\!2\biggl({4m_t^2-M_h^2\over s}\biggr)
+{2\over x_h}
{(4m_t^2-M_h^2)(2m_t^2-M_h^2)\over s^2
}
+{8m_t^2\over s x_h}\biggl]\ln\biggl({x_h+
{\hat\beta}\over x_h-{\hat \beta}}\biggr)
\biggr\}\nonumber\,\,\,\,,
\end{eqnarray}
where $x_h\!=\!2 E_h/\sqrt{s}$, $E_h$ is the Higgs boson energy in the
$q\bar q$ center-of-mass frame, $x_h^{min}\!=\!2 M_h/\sqrt{s}$,
$x_h^{max}=1-4m_t^2/s +M_h^2/s$, and we have introduced:
\begin{equation}
\label{eq:beta_xh}
{\hat \beta}=\biggl\{{[x_h^2-(x_h^{min})^2][x_h^{max}-x_h]
\over x_h^{max}-x_h+{4 m_t^2/s}}\biggr\}^{1/2}\,\,\,\,.
\end{equation}
Moreover, we have defined the Yukawa coupling of the top quark to be
$g_t\!=m_t/v$, where $v\!=\!(G_F\sqrt{2})^{-1/2}$ is the vacuum
expectation value of the SM Higgs boson, given in terms of the Fermi
constant $G_F$.

\begin{figure}[hbt]
\centering
\epsfxsize=4.3in
\leavevmode\epsffile{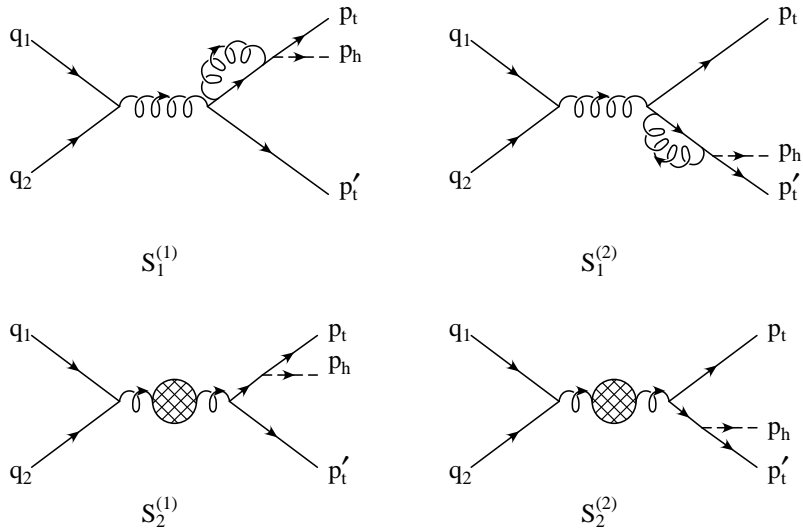}
\caption[]{${\cal O}(\alpha_s)$ virtual corrections: self-energy 
diagrams $S_1^{(1,2)}$ and $S_2^{(1,2)}$.}
\label{fig:selfenergies}
\end{figure}

The NLO QCD contribution, $\delta{\hat\sigma}_{\sss
  NLO}^{ij}(x_1,x_2,\mu)$, contains both virtual and real ${\cal
  O}(\alpha_s)$ corrections to the lowest-order cross section and can
be written as the sum of two terms:
\begin{eqnarray}
\label{eq:delta_sigmahat}
\delta {\hat \sigma}_{\sss NLO}^{ij}(x_1,x_2,\mu)&= &
\int d(PS_3) \overline{\sum}|{\cal A}_{virt}(i j \rightarrow t\bar th)|^2+
\int d(PS_4)\overline{\sum}|{\cal A}_{real}(ij\rightarrow t\bar th+g)|^2
\nonumber \\
&\equiv&\hat{\sigma}^{ij}_{virt}(x_1,x_2,\mu)+
\hat{\sigma}^{ij}_{real}(x_1,x_2,\mu) \,\,\,,
\end{eqnarray}
where $|{\cal A}_{virt}(ij \rightarrow t\bar th )|^2$ and $|{\cal
  A}_{real}(ij \rightarrow t\bar th+g)|^2$ are respectively the
squared matrix elements for the ${\cal O}(\alpha_s^3)$ $ij\rightarrow
t\bar th$ and $ij\rightarrow t\bar t h+g$ processes, and
$\overline{\sum}$ indicates that they have been averaged over the
initial-state degrees of freedom and summed over the final-state ones.
Moreover, $d(PS_3)$ and $d(PS_4)$ denote the integration over the
corresponding three and four-particle phase spaces respectively.  The
first term in Eq.~(\ref{eq:delta_sigmahat}) represents the
contribution of the virtual gluon corrections, while the second one is
due to the real gluon emission. For the $q\bar q\rightarrow t\bar th$
sub-process, examples of ${\cal O}(\alpha_s)$ virtual and real
corrections are illustrated in
Figs.~\ref{fig:selfenergies}-\ref{fig:real} and their structure is
separately explained in Secs.~\ref{sec:virtual} and \ref{sec:real}.

Finally, we observe that, in order to assure the renormalization scale
independence of the total cross section at ${\cal O}(\alpha_s^3)$,
$f_{\sss NLO}^{ij}(x_1,x_2,\mu)$ in Eq.(\ref{eq:sigmahat_nlo}) must be
of the form:
\begin{equation}
f_{\sss NLO}^{ij}(x_1,x_2,\mu)=f_1^{ij}(x_1,x_2)+
\tilde{f}_1^{ij}(x_1,x_2)\ln\left(\frac{\mu^2}{s}\right)\,\,\,,
\end{equation}
with $\tilde{f}_1^{ij}(x_1,x_2)$ given by:
\begin{eqnarray}
\label{eq:mudep_coeff}
\tilde{f}_1^{ij}(x_1,x_2)&=&
2\,\left\{4\pi b_0 f^{ij}_{\sss LO}(x_1,x_2)-
\sum_k\left[\int_\rho^1 dz_1 P_{ik}(z_1)f^{kj}_{\sss LO}(x_1 z_1,x_2)
\right. \right.\\
&+&\left.\left.\int_\rho^1 dz_2 P_{kj}(z_2)f^{ik}_{\sss LO}(x_1,x_2z_2) 
\right]\right\}
\,\,\,,\nonumber
\end{eqnarray}
where $\rho\!=\!(2m_t+M_h)^2/s$, $P_{ij}(z)$ denotes the lowest-order
Altarelli-Parisi splitting function \cite{Altarelli:1977zs} of parton
$i$ into parton $j$, when $j$ carries a fraction $z$ of the momentum
of parton $i$, (see e.g.  Sec.~\ref{subsubsec:two_cutoff_hard}), and
$b_0$ is determined by the one-loop renormalization group evolution of
the strong coupling constant $\alpha_s$:
\begin{equation}
\frac{d\alpha_s(\mu)}{d\ln(\mu^2)}=-b_0\alpha_s^2+{\cal
  O}(\alpha_s^3)\,\,\,\,\,,
\,\,\,\,\,
b_0=\frac{1}{4\pi}\left(\frac{11}{3}N-\frac{2}{3}n_{lf}\right)\,\,\,,
\end{equation}
with $N=3$, the number of colors, and $n_{lf}\!=\!5$, the number of
light flavors. The origin of the terms in Eq.~(\ref{eq:mudep_coeff})
will become manifest in Secs.~\ref{sec:virtual} and \ref{sec:real},
when we describe in detail the calculation of both virtual and real
${\cal O}(\alpha_s)$ corrections to $q \bar q\rightarrow t\bar th$.

\section{Virtual Corrections}
\label{sec:virtual}
\begin{figure}[hbtp]
\vspace{-1.7truecm} 
\centering \epsfxsize=4.3in
\leavevmode\epsffile{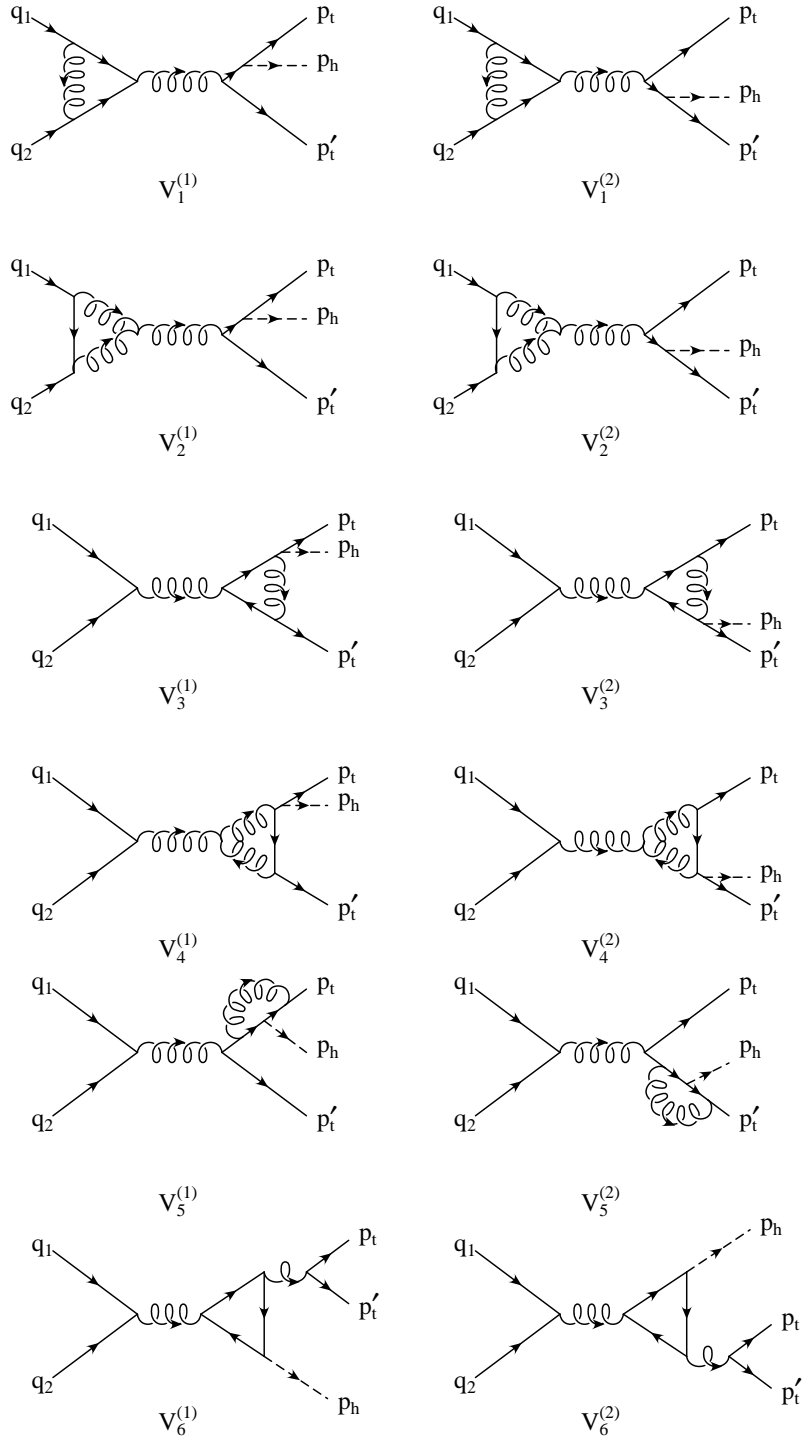}
\caption[]{${\cal O}(\alpha_s)$ virtual corrections: vertex 
diagrams $V_1^{(1,2)}$-$V_6^{(1,2)}$.}
\label{fig:vertices}
\end{figure}
\begin{figure}[hbtp]
\centering
\epsfxsize=4.3in
\leavevmode\epsffile{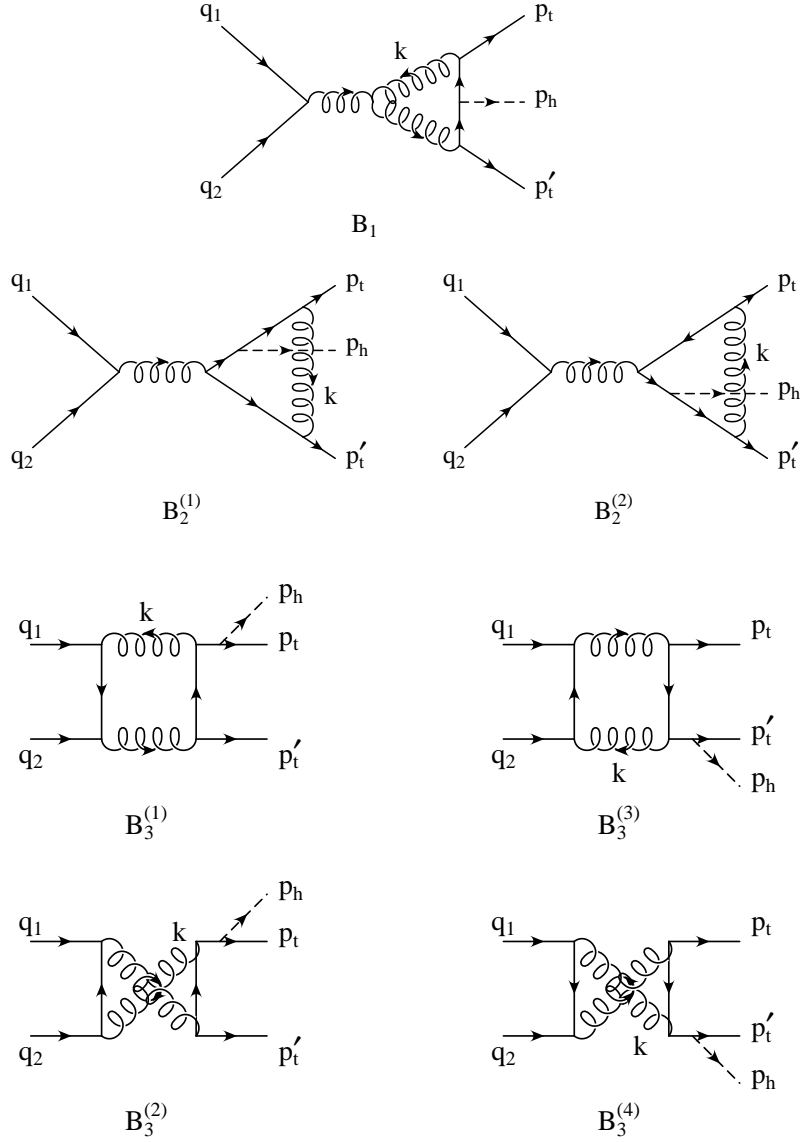}
\caption[]{${\cal O}(\alpha_s)$ virtual corrections: box 
diagrams $B_1$, $B_2^{(1,2)}$ and $B_3^{(1-4)}$.}
\label{fig:boxes}
\end{figure}
\begin{figure}[hbt]
\centering
\epsfxsize=4.3in
\leavevmode\epsffile{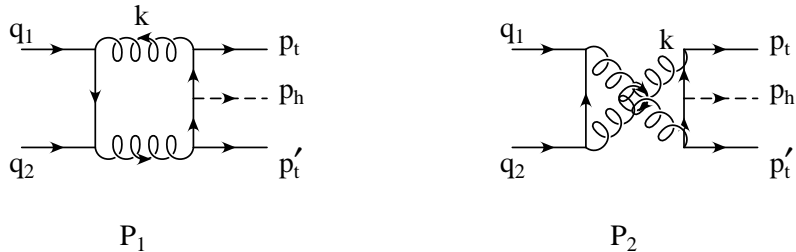}
\caption[]{${\cal O}(\alpha_s)$ virtual corrections: pentagon 
diagrams $P_1$ and $P_2$.}
\label{fig:pentagons}
\end{figure}
\begin{figure}[hbt]
\centering
\epsfxsize=4.3in
\leavevmode\epsffile{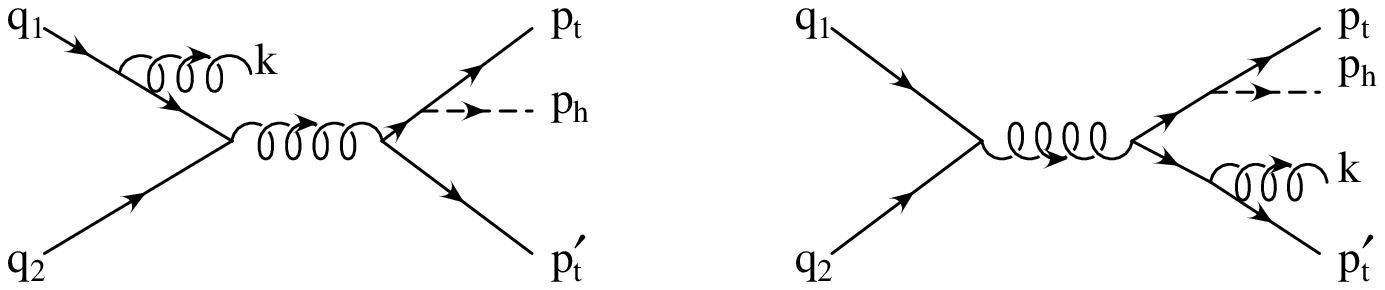}
\caption[]{${\cal O}(\alpha_s)$ real corrections: examples of initial
  and final state real gluon emission.}
\label{fig:real}
\end{figure}
The ${\cal O}(\alpha_s)$ virtual corrections to the tree-level $ q
{\overline q}\rightarrow t {\overline t} h$ process consist of
self-energy, vertex, box, and pentagon diagrams which are shown in
Figs.~\ref{fig:selfenergies}-\ref{fig:pentagons}.  We assign incoming
and outgoing momenta according to the following notation,
\begin{equation}
q(q_1)+{\overline q}(q_2) \rightarrow t(p_t) +{\overline t}(p_t^\prime)
+h(p_h)\,\,\,,
\end{equation}
where the momentum flow is illustrated in
Figs.~\ref{fig:selfenergies}-\ref{fig:pentagons}. If we denote by
${\cal A}_{D_i}$ the amplitude associated with each virtual diagram
$D_i$, the ${\cal O}(\alpha_s^3)$ virtual amplitude squared can then
be written as:
\begin{equation}
\overline{\sum}|{\cal A}_{virt}|^2=
\sum_i\overline{\sum}\left({\cal A}_{\sss LO}^d {\cal A}_{D_i}^*+
{\cal A}_{\sss LO}^{d*} {\cal A}_{D_i}\right)=
\sum_i\overline{\sum}2\,{\cal R}e\left({\cal A}_{\sss LO}^d 
{\cal A}_{D_i}^*\right)\,\,\,,
\label{eq:amp2_virt_gen}
\end{equation}
where the index $i$ runs over the set of all virtual diagrams, and
${\cal A}_{\sss LO}^d$ denotes the tree-level amplitude for $q\bar
q\rightarrow t\bar t h$ calculated in $d\!=\!4-2\epsilon$ dimensions.
The lowest order amplitude ${\cal A}_{\sss LO}^d$ must be computed to
${\cal O}(\epsilon^2)$ in order to properly account for both the
singular and finite contributions generated by the interference of
${\cal A}_{\sss LO}^d$ with the single and double poles present in the
virtual amplitudes ${\cal A}_{D_i}$.  In what follows, we denote by
${\cal A}_{\sss LO}$ the lowest order amplitude to ${\cal
  O}(\epsilon^0)$, i.e. calculated in $d\!=\!4$ dimensions.  Also, in
the following sections, the contribution of a given diagram or set of
diagrams to $\hat{\sigma}_{virt}^{q\bar q}$ is always to be understood
as the contribution of the corresponding term in the sum in
Eq.~(\ref{eq:amp2_virt_gen}).

The calculation of the virtual diagrams has been performed using
dimensional regularization, always in $d\!=\!4-2\epsilon$ dimensions.
The diagrams have been evaluated using \emph{FORM}
\cite{Vermaseren:2000nd} and \emph{Maple}, and all tensor integrals
have been reduced to linear combinations of a fundamental set of
scalar one-loop integrals using standard techniques
\cite{'tHooft:1979xw}.  The scalar integrals which give rise to either
ultraviolet (UV) or infrared (IR) singularities have been computed
analytically, while finite scalar integrals have been evaluated using
standard packages \cite{vanOldenborgh:1990wn}.

Self-energy and vertex diagrams contain both IR and UV divergences.
The UV divergences are renormalized by introducing a suitable set of
counterterms. Since the cross section is a renormalization group
invariant, we only need to renormalize the wave function of the
external fields, the top-quark mass, and the coupling constants. We discuss
the renormalization of the UV singularities of the virtual cross
section in Sec.~\ref{subsec:virtual_uv}.

Box and pentagon diagrams are ultraviolet finite, but have infrared
singularities. The IR poles in the ${\cal O}(\alpha_s)$ virtual
corrections are eventually canceled by analogous singularities in the
${\cal O}(\alpha_s)$ real corrections to the tree-level cross section.
We discuss the structure of the IR singularities of the virtual cross
section in Sec.~\ref{subsec:virtual_ir}. The structure of the IR
singularities of the real cross section will be the subject of
Secs.~\ref{subsec:two_cutoff} and \ref{subsec:one_cutoff}.

The calculation of many of the box scalar integrals and in particular
of the pentagon scalar integrals are extremely laborious, due to the
large number of massive particles present in the final state and in
the loop. We have evaluated the necessary pentagon scalar integrals
(one for diagram $P_1$ and one for diagram $P_2$), using the method of
Ref.~\cite{Bern:1993em}, which allows the reduction of a scalar
five-point function to a sum of five scalar four-point functions, plus
terms of $O(\epsilon)$ which can be neglected. Since this is a crucial
ingredient of this calculation, we will explain in detail in
Appendix~\ref{sec:scalar_pentagon} how the method of
Ref.~\cite{Bern:1993em} has been applied to our case. The IR-divergent
box scalar integrals are also collected in
Appendix~\ref{sec:scalar_boxes}.

\subsection{UV singularities and counterterms}
\label{subsec:virtual_uv}

The UV singularities of the ${\cal O}(\alpha_s^3)$ total cross section
originate from self-energy and vertex virtual corrections.  These
singularities are renormalized by introducing counterterms for the
wave function of the external fields ($\delta Z_2^{(q)}$, $\delta
Z_2^{(t)}$), the top-quark mass ($\delta m_t$), and the coupling constants
($\delta g_t$, $\delta Z_{\alpha_s}$).  If we denote by
$\Delta_{UV}(S_i^{(1,2)})$ and $\Delta_{UV}(V_i^{(1,2)})$ the UV-divergent
contribution of each self-energy ($S_i^{(1,2)}$) or vertex diagram
($V_i^{(1,2)}$) to the virtual amplitude squared (see
Eq.~(\ref{eq:amp2_virt_gen})), we can write the UV-singular part of
the total virtual amplitude squared as:
\begin{eqnarray}
\label{eq:amp2_virt_uv}
\overline{\sum}|{\cal A}_{virt}^{\sss UV}|^2&=& \overline{\sum} 
|{\cal A}_{\sss LO}|^2
\,\left\{\sum_{i=1}^2\Delta_{UV}(S_i^{(1)}+S_i^{(2)})+
\,\sum_{i=1}^6\Delta_{UV}(V_i^{(1)}+V_i^{(2)})\right.\\
&+&\left.2\,\left[\,\left(\delta Z_2^{(q)}\right)_{UV}+
\left(\delta Z_2^{(t)}\right)_{UV}+\frac{\delta m_t}{m_t}+
\delta Z_{\alpha_s}\right]\right\}\,\,\,.\nonumber
\end{eqnarray}
As described earlier, we denote by $|{\cal A}_{\sss LO}|^2$ the matrix
element squared of the tree-level amplitude for $q\bar q\rightarrow
t\bar t h$, computed in $d=4$ dimensions.  We also notice that, in
writing Eq.~(\ref{eq:amp2_virt_uv}), we have included in the top-quark
self-energy the top-mass counterterm, and we have used the fact that
the Yukawa-coupling counterterm coincides with the top-mass
counterterm.

The UV-divergent contributions due to the individual diagrams are
explicitly given by:
\begin{eqnarray}
\label{eq:virtual_uv}
\Delta_{UV}\left(S_1^{(1)}+S_1^{(2)}\right)&=& 
\frac{\alpha_s}{2\pi}{\cal N}_t\left(\frac{N}{2}-\frac{1}{2N}\right)
\biggl(-{1\over \epsilon_{\sss UV}}\biggr)\,\,\,,\nonumber\\
\Delta_{UV}\left(S_2^{(1)}+S_2^{(2)}\right)&=&
\frac{\alpha_s}{2\pi}\biggl[ {\cal N}_s
\left(\frac{5}{3}N-\frac{2}{3}n_{lf}\right)-{\cal N}_t\frac{2}{3}\biggr]
\biggl({1\over \epsilon_{\sss UV}}\biggr)\,\,\,,
\nonumber\\
\Delta_{UV}\left(V_1^{(1)}+V_1^{(2)}\right)&=& \frac{\alpha_s}{2\pi}
{\cal N}_s\left(-{1\over 2 N}\right)
\biggl({1\over \epsilon_{\sss UV}}\biggr)\,\,\,,\nonumber\\
\Delta_{UV}\left(V_2^{(1)}+V_2^{(2)}\right)&=& \frac{\alpha_s}{2\pi}
{\cal N}_s\left(\frac{N}{2}\right)
\biggl({3\over \epsilon_{\sss UV}}\biggr)\,\,\,,\\
\Delta_{UV}\left(V_3^{(1)}+V_3^{(2)}\right)&=& \frac{\alpha_s}{2\pi}
{\cal N}_t\left(-{1\over 2N}\right)
\biggl({1\over \epsilon_{\sss UV}}\biggr)\,\,\,,\nonumber\\
\Delta_{UV}\left(V_4^{(1)}+V_4^{(2)}\right)&=& \frac{\alpha_s}{2\pi}
{\cal N}_t\left(\frac{N}{2}\right)
\biggl({3\over \epsilon_{\sss UV}}\biggr)\,\,\,,\nonumber\\
\Delta_{UV}\left(V_5^{(1)}+V_5^{(2)}\right)&=& \frac{\alpha_s}{2\pi}
{\cal N}_t\left(\frac{N}{2}-\frac{1}{2N}\right)
\biggl({4\over \epsilon_{\sss UV}}\biggr)\,\,\,,\nonumber\\
\Delta_{UV}\left(V_6^{(1)}+V_6^{(2)}\right)&=& 0\,\,\,,\nonumber
\end{eqnarray}
where ${\cal N}_s$ and ${\cal N}_t$ are standard normalization factors
defined as:
\begin{equation}
\label{eq:nsnt}
{\cal N}_s=
\biggl({4 \pi \mu^2\over s}\biggr)^\epsilon
\Gamma(1+\epsilon)\,\,\,\,,\,\,\,\,
{\cal N}_t=
\biggl({4 \pi \mu^2\over m_t^2}\biggr)^\epsilon
\Gamma(1+\epsilon)\,\,\,\,.
\end{equation}

Moreover we define the needed counterterms according to the following
convention.  For the external fields, we fix the wave-function
renormalization constants of the external fields ($Z_2^{(i)}=1+\delta
Z_2^{(i)}$, $i\!=\!q,t$) using on-shell subtraction, i.e.:
\begin{eqnarray}
\label{eq:z2_ct}
\left(\delta Z_2^{(q)}\right)_{UV}
&=&-\biggl({\alpha_s\over 4 \pi}\biggr){\cal N}_s
\biggl({N\over 2}-{1\over 2 N}\biggr)
 \biggl({1\over \epsilon_{\sss UV}}\biggr)\,\,\,,\\
\left(\delta Z_2^{(t)}\right)_{UV}
&=&-\biggl({\alpha_s\over 4 \pi}\biggr){\cal N}_t
\biggl({N\over 2}-{1\over 2 N}\biggr)
 \biggl({1\over \epsilon_{\sss UV}}+4\biggr)\,\,\,.\nonumber
\end{eqnarray}
We notice that both $\delta Z_2^{(q)}$ and $\delta Z_2^{(t)}$, as well
as some of the vertex corrections ($V_1^{(1,2)}$ and $V_2^{(1,2)}$),
have also IR singularities. In this section we limit the discussion to
the UV singularities only, while the IR structure of these terms will
be given explicitly in Sec.~\ref{subsec:virtual_ir}.

We define the subtraction condition for the top-quark mass $m_t$ in
such a way that $m_t$ is the pole mass, in which case the top-mass
counterterm is given by:
\begin{equation}
\label{eq:mt_ct}
{\delta m_t\over m_t}=-\biggl({\alpha_s\over 4 \pi}\biggr){\cal N}_t
\biggl({N\over 2}-{1\over 2 N}\biggr)
\biggl({3\over \epsilon_{\sss UV}}+4\biggr)\,\,\,.
\end{equation}
This counterterm has to be used twice: to renormalize the top-quark
mass, in diagrams $S_1^{(1)}$ and $S_1^{(2)}$, and to renormalize the
top-quark Yukawa coupling. As we already noted,
$\Delta_{UV}(S_1^{(1)}+S_1^{(2)})$ in Eq.~(\ref{eq:virtual_uv})
already includes the top-mass counterterm.

Finally, for the renormalization of $\alpha_s$ we use the
$\overline{MS}$ scheme, modified to decouple the top quark
\cite{Collins:1978wz}.  The first $n_{lf}$ light flavors are
subtracted using the $\overline{MS}$ scheme, while the divergences
associated with the top-quark loop are subtracted at zero momentum:
\begin{equation}
\label{eq:alphas_ct}
\delta Z_{\alpha_s}=\biggl({\alpha_s\over 4 \pi}\biggr)({4 \pi})^\epsilon
\Gamma(1+\epsilon)
\biggl[ \biggl(\frac{2}{3}n_{lf} - {11\over 3} N\biggr)
{1\over \epsilon_{\sss UV}}+
\frac{2}{3} \biggl( {1\over \epsilon_{\sss UV}}+
\ln\biggl(\frac{\mu^2}{m_t^2}\biggr)
\biggr)\biggr]\,\,\,,
\end{equation}
such that, in this scheme, the renormalized strong coupling constant
$\alpha_s$ evolves with $n_{lf}=5$ light flavors.

It is easy to verify that the sum of all the UV-singular
contributions as given in
Eq.~(\ref{eq:amp2_virt_uv}) is finite. We also notice that the left
over renormalization scale dependence, due to the mismatch between the
renormalization scale dependence of $\Delta_{UV}(S_2)$ and
$\delta(Z_{\alpha_s})$, is given by:
\begin{equation}
\overline{\sum}|{\cal A}_{\sss LO}|^2 {\alpha_s(\mu)\over 2 \pi}
\left(-\frac{2}{3} n_{lf}+\frac{11}{3}N\right) 
\ln\left(\frac{\mu^2}{s}\right)\,\,\,,
\end{equation}
and corresponds exactly to the first term of
Eq.~(\ref{eq:mudep_coeff}), as predicted by renormalization group
arguments.

\subsection{IR singularities}
\label{subsec:virtual_ir}
This section describes the structure of the IR singularities
originating from the ${\cal O}(\alpha_s)$ virtual corrections.  The
virtual IR singularities come from the following set of diagrams:
vertex diagrams $V_1^{(1,2)}$ and $V_2^{(1,2)}$, box diagrams
$B_2^{(1,2)}$, box diagrams $B_3^{(1-4)}$, pentagon diagrams $P_1$ and
$P_2$, and from the wave function renormalization of the external
fields, $\delta Z_2^{(q)}$ and $\delta Z_2^{(t)}$. The IR-singular
part of the total virtual amplitude squared is then of the form:
\begin{eqnarray}
\label{eq:sigmavirt_ir}
\overline{\sum}|{\cal A}_{virt}^{\sss IR}|^2 
&=&
\overline{\sum}|{\cal A}_{\sss LO}|^2\left\{
\Delta_{IR}\left(V_1^{(1)}+V_1^{(2)}\right)+
\Delta_{IR}\left(V_2^{(1)}+V_2^{(2)}\right)+
\left(\delta Z_2^{(q)}\right)_{IR}+
\left(\delta Z_2^{(q)}\right)_{IR}
\right.\nonumber\\
&&\!\!\!\!\!\!\!\!\!\!\!\!\left.+\,
\Delta_{IR}\left(B_2^{(1)}+B_2^{(2)}\right)+
\Delta_{IR}\left(B_3^{(1)}+B_3^{(3)}+P_1\right)+
\Delta_{IR}\left(B_3^{(2)}+B_3^{(4)}+P_1\right)\right\}\,\,\, ,
\end{eqnarray}
where, as before, $|{\cal A}_{\sss LO}|^2$ denotes the matrix element
squared of the tree-level amplitude for $q\bar q\rightarrow t\bar t
h$, in $d\!=\!4$ dimensions. The IR-divergent contributions of the
various diagrams to the virtual amplitude squared are given in the
following:
\begin{eqnarray}
\label{eq:virtual_ir}
\Delta_{IR}\left(V_1^{(1)}+V_1^{(2)}\right)&=& 
\left(\frac{\alpha_s}{2\pi}\right){\cal N}_s
\left(-\frac{1}{2N}\right)
\left(-\frac{2}{\epsilon_{\sss IR}^2}-\frac{4}{\epsilon_{\sss IR}}\right)
\,\,\,,\nonumber\\
\Delta_{IR}\left(V_2^{(1)}+V_2^{(2)}\right)&=&
\left(\frac{\alpha_s}{2\pi}\right){\cal N}_s
\left(\frac{N}{2}\right)
\left(-\frac{4}{\epsilon_{\sss IR}}\right)\,\,\,,\nonumber\\
\left(\delta Z_2^{(q)}\right)_{IR}&=&
\left(\frac{\alpha_s}{2\pi}\right){\cal N}_s
\left(\frac{N}{2}-\frac{1}{2N}\right)
\left(\frac{1}{\epsilon_{\sss IR}}\right)\,\,\,,\\
\left(\delta Z_2^{(t)}\right)_{IR}&=&
\left(\frac{\alpha_s}{2\pi}\right){\cal N}_t
\left(\frac{N}{2}-\frac{1}{2N}\right)
\left(-\frac{2}{\epsilon_{\sss IR}}\right)\,\,\,,\nonumber\\
\Delta_{IR}\left(B_2^{(1)}+B_2^{(2)}\right)&=&
\left(\frac{\alpha_s}{2\pi}\right){\cal N}_t
\left(-\frac{1}{N}\right)
\left(\frac{1}{\epsilon_{\sss IR}}
\frac{s_{t\bar t}}{(2 m_t^2+s_{t\bar t})
\beta_{t\bar t}}\Lambda_{t\bar t}\right)\,\,\,,\nonumber\\
\Delta_{IR}\left(B_3^{(1)}+B_3^{(3)}+P_1\right)&=& 
\left(\frac{\alpha_s}{2\pi}\right){\cal N}_t
\left(\frac{N}{2}-\frac{1}{N}\right)
\left[-\frac{2}{\epsilon_{\sss IR}^2}+\frac{2}{\epsilon_{\sss IR}}\left(
\ln\left(\frac{s_{qt}}{m_t^2}\right)+
\ln\left(\frac{s_{\bar q\bar t}}{m_t^2}\right)
\right)\right]\,\,\,,\nonumber\\ 
\Delta_{IR}\left(B_3^{(2)}+B_3^{(4)}+P_2\right)&=&
\left(\frac{\alpha_s}{2\pi}\right)
{\cal N}_t \left(-\frac{1}{N}\right)
\left[\frac{2}{\epsilon^2_{\sss IR}}-\frac{2}{\epsilon_{\sss IR}}\left(
\ln\left(\frac{s_{q\bar t}}{m_t^2}\right)+
\ln\left(\frac{s_{\bar qt}}{m_t^2}\right)
\right)\right]\,\,\,, \nonumber
\end{eqnarray}
where ${\cal N}_s$ and ${\cal N}_t$ are given in Eq.~(\ref{eq:nsnt}).
Moreover, we have introduced the following kinematic invariants: 
\begin{equation}
\label{eq:ir_invariants}
s\!=s_{q\bar q}=\!2q_1\cdot q_2\,\,\,,\,\,\,
s_{t\bar t}\!=\!2p_t\cdot p_t^\prime\,\,\,,\,\,\,
s_{qt}\!=\!2q_1\cdot p_t\,\,\,,\,\,\,
s_{q\bar t}\!=\!2q_1\cdot p_t^\prime\,\,\,,\,\,\,
s_{\bar qt}\!=\!2q_2\cdot p_t\,\,\,,\,\,\,
s_{\bar q\bar t}\!=\!2q_2\cdot p_t^\prime\,\,\,,\,\,\,
\end{equation}
and we have defined
\begin{eqnarray}
\label{eq:betadef}
\beta_{t\bar t} &=&\sqrt{1-{4 m_t^2\over (p_t
        +p_t^\prime)^2}}\,\,\,, \\
\Lambda_{t\bar t} &=& \ln\biggl({1+\beta_{t\bar t}
\over 1 -\beta_{t\bar t}}\biggr)\,\,\,.\nonumber
\end{eqnarray}
Substituting the explicit expression for the IR-divergent
contributions given in Eq.~(\ref{eq:virtual_ir}) into
Eq.~(\ref{eq:sigmavirt_ir}) yields:
\begin{equation}
\overline{\sum}|{\cal A}_{virt}^{\sss IR}|^2 
=  
\left(\frac{\alpha_s}{2\pi}\right){\cal N}_t\,
\overline{\sum}|{\cal A}_{\sss LO}|^2
\left\{
\frac{X_{-2}^{virt}}{\epsilon_{\sss IR}^2}+
\frac{X_{-1}^{virt}}{\epsilon_{\sss IR}}
+\delta_{virt}^{\sss IR}\right\}\,\,\,,
\end{equation}
where
\begin{eqnarray}
\label{eq:sigma_ir_poles}
X_{-2}^{virt}&=& -\left(N-\frac{1}{N}\right)\,\,\, ,\\
X_{-1}^{virt}&=& N\,\left[-\frac{5}{2}
+\ln\left(\frac{s_{q t}}{m_t^2}\right)
+\ln\left(\frac{s_{\bar q\bar t}}{m_t^2}\right)\right]\nonumber\\
&+&\frac{1}{N}\,\left[-\ln\left(\frac{s}{m_t^2}\right)+\frac{5}{2}
-\frac{s_{t\bar t}}{(2 m_t^2+s_{t\bar t})\beta_{t\bar t}}
\Lambda_{t\bar t}-2\ln\left(\frac{s_{q t}s_{\bar q\bar t}}
{s_{q\bar t}s_{\bar qt}}\right)\right]\,\,\,,\nonumber
\end{eqnarray}
while $\delta_{virt}^{IR}$ is a finite term that derives from having
factored out a common factor ${\cal N}_t$, and is given by:
\begin{equation}
\delta_{virt}^{\sss IR}=\left(N-\frac{1}{N}\right)
\left[\frac{3}{2}\ln\left(\frac{s}{m_t^2}\right)\right]+
\frac{1}{N}\left[\frac{1}{2}\ln^2\left(\frac{s}{m_t^2}\right)\right]\,\,\,.
\end{equation}
In Secs.~\ref{subsubsec:two_cutoff_soft} and
\ref{subsubsec:one_cutoff_ir} we will show how the IR singularities of
the real cross section exactly cancel the IR poles of the virtual
cross section (see
Eqs.~(\ref{eq:sigma_soft_total})-(\ref{eq:sigma_soft_poles}) and
Eqs.~(\ref{eq:a2_soft_coll})-(\ref{eq:sigma_soft_coll_coeff})), as
predicted by the Bloch-Nordsieck \cite{Bloch:1937pw} and
Kinoshita-Lee-Nauenberg \cite{Kinoshita:1962ur,Lee:1964is} theorems.

\section{Real Corrections}
\label{sec:real}
The ${\cal O}(\alpha_s)$ corrections to $q\bar q\rightarrow t\bar th$
due to real gluon emission (see Fig.~\ref{fig:real}) give origin to IR
singularities which cancel exactly the analogous singularities present
in the ${\cal O}(\alpha_s)$ virtual corrections (see
Sec.~\ref{subsec:virtual_ir}). These singularities can be either of
\emph{soft} or \emph{collinear} nature and can be conveniently
isolated by \emph{slicing} the $q\bar q\rightarrow t\bar th+g$ phase
space into different regions defined by suitable cutoffs, a method
which goes under the general name of \emph{Phase Space Slicing} (PSS).
The dependence on the arbitrary cutoff(s) introduced in the process is
not physical, and, in fact, cancels at the level of the total real
gluon emission hadronic cross section, i.e. in $\sigma_{real}$, the
real part of $\sigma_{\sss NLO}$ . This constitutes an important check
of the calculation.

We have calculated the cross section for the process
\begin{equation}
q (q_1)+{\overline q}(q_2) \rightarrow
t (p_t)+{\overline t}(p_t^\prime) + h(p_h)+g(k)
\end{equation}
using two different implementations of the PSS method which we call
the \emph{two-cutoff} and \emph{one-cutoff} method respectively,
depending on the number of cutoffs introduced.  The \emph{two-cutoff}
implementation of the PSS method has been originally developed to
study QCD corrections to dihadron production \cite{bergmann} and has
since then been applied to a variety of processes. A nice review has
recently appeared \cite{Harris:2001sx} to which we refer for more
extensive references and details.  The \emph{one-cutoff} PSS method
has been developed for massless quarks in
Ref.~\cite{Giele:1992vf,Giele:1993dj} and extended to the case of
massive quarks in Ref.~\cite{Keller:1999tf}.

In the next two sections we explain in detail how we have applied the
PSS method to our case, using the \emph{two-cutoff} implementation in
Sec.~\ref{subsec:two_cutoff} and the \emph{one-cutoff} implementation
in Sec.~\ref{subsec:one_cutoff}. The results for $\sigma_{real}$
obtained using PSS with one or two cutoffs agree within the
statistical errors.  In spite of the fact that both methods are
realizations of the general idea of phase space slicing, they have
very different characteristics and finding agreement between the two
represents an important check of our calculation.

\subsection{Phase Space Slicing method with two cutoffs}
\label{subsec:two_cutoff}

The general implementation of the PSS method using two cutoffs
proceeds in two steps. First, by introducing an arbitrary small
\emph{soft} cutoff $\delta_s$ we separate the overall integration of
the $q\bar q\rightarrow t\bar th+g$ phase space into two regions,
according to whether the energy of the gluon is \emph{soft}, i.e.
$E_g\le\delta_s\sqrt{s}/2$, or \emph{hard}, i.e.
$E_g>\delta_s\sqrt{s}/2$. The partonic real cross section of
Eq.~(\ref{eq:delta_sigmahat}) can then be written as:
\begin{equation}
\label{eq:sigma_real_two_cutoff}
\hat{\sigma}^{q\bar q}_{real} =
\hat{\sigma}_{soft}+\hat{\sigma}_{hard}\,\,\,,
\end{equation}
where $\hat\sigma_{soft}$ is obtained by integrating over the
\emph{soft} region of the gluon phase space, and contains all the IR
soft divergences of $\hat\sigma_{real}^{q\bar q}$.  To isolate the
remaining collinear divergences from $\hat\sigma_{hard}$, we further
split the integration over the hard gluon phase space according to
whether the gluon is ($\hat\sigma_{hard/coll}$) or is not
($\hat\sigma_{hard/non-coll}$) emitted within an angle $\theta$ from
the initial-state massless quarks such that $(1-\cos\theta)<\delta_c$,
for an arbitrary small \emph{collinear} cutoff $\delta_c$:
\begin{equation}
\label{eq:hardsplit}
\hat{\sigma}_{hard}=\hat{\sigma}_{hard/coll}+
\hat{\sigma}_{hard/non-coll}\,\,\,.
\end{equation}
The hard non-collinear part of the real cross section,
$\hat{\sigma}_{hard/non-coll}$, is finite and can be computed
numerically, using standard Monte-Carlo techniques.  In the soft and
collinear regions, the integration over the phase space of the emitted
gluon can be performed analytically, thus allowing us to isolate the
IR collinear divergences of $\hat{\sigma}_{real}^{q \bar q}$. More
details on the calculation of $\hat{\sigma}_{soft}$ and
$\hat\sigma_{hard}$ are given in Sec.~\ref{subsubsec:two_cutoff_soft}
and Sec.~\ref{subsubsec:two_cutoff_hard}, respectively.  The cross
sections describing soft, collinear and IR-finite gluon radiation
depend on the two arbitrary parameters, $\delta_s$ and $\delta_c$.
However, in the real hadronic cross section $\sigma_{real}$, after
mass factorization, the dependence on these arbitrary cutoffs cancels,
as will be explicitly shown in Sec.~\ref{sec:total}.

\subsubsection{Soft gluon emission}
\label{subsubsec:two_cutoff_soft} 

The soft region of the $q\bar q\rightarrow t\bar t h+g$ phase space
is defined by requiring that the energy of the gluon satisfies:
\begin{equation}
E_g <\delta_s {\sqrt{s}\over 2}\,\,\,,
\end{equation}
for an arbitrary small value of the \emph{soft} cutoff $\delta_s$.  In
the limit when the energy of the gluon becomes small, i.e. in the
\emph{soft limit}, the matrix element squared for the real gluon
emission, $\overline{\sum} |{\cal A}_{real}|^2$, assumes a very simple
form, i.e.  it factorizes into the Born matrix element squared times
an eikonal factor $\Phi_{eik}$:
\begin{equation}
\label{eq:m2_soft_lim}
\overline{\sum}|{\cal A}_{real}(q\bar q\rightarrow t\bar t h+g)|^2 
\stackrel{soft}{\longrightarrow}
 (4 \pi \alpha_s) \overline{\sum}|{\cal A}_{\sss LO}|^2\,\Phi_{eik}\,\,\,,
\end{equation}
where the eikonal factor is given by:
\begin{eqnarray}
\label{eq:eik_factor}
\Phi_{eik}&=&
\frac{N}{2}\left[
-\frac{m_t^2}{(p_t\!\cdot\! k)^2}
-\frac{m_t^2}{(p_t^\prime\!\cdot\! k)^2}
+\frac{s_{qt}}{(q_1\!\cdot\! k)(p_t\!\cdot\! k)}
+\frac{s_{\bar q\bar t}}{(q_2\!\cdot\! k)(p_t^\prime\!\cdot\! k)}\right]
\\
&& +
\frac{1}{2N} \left[
\frac{m_t^2}{(p_t\!\cdot\! k)^2}
+\frac{m_t^2}{(p_t^\prime\!\cdot\! k)^2}
-\frac{s}{(q_1\!\cdot\! k)(q_2\!\cdot\! k)}
-\frac{s_{t\bar t}}{(p_t\!\cdot\! k)(p_t^\prime\!\cdot\! k)}
\right. \nonumber \\
&&\left.
+2\left( -\frac{s_{qt}}{(q_1\!\cdot\! k)(p_t\!\cdot\! k)}
+\frac{s_{q\bar t}}{(q_1\!\cdot\! k)(p_t^\prime\!\cdot\! k)}
+\frac{s_{\bar qt}}{(q_2\!\cdot\! k)(p_t\!\cdot\! k)}
-\frac{s_{\bar q\bar t}}{(q_2\!\cdot\! k)(p_t^\prime\!\cdot\! k)}
\right)\right]\,\,\,.\nonumber
\end{eqnarray}
Moreover, in the soft region the $q\bar q\rightarrow t\bar th+g$ 
phase space also factorizes as:
\begin{eqnarray}
\label{eq:ps_soft_lim}
d(PS_4)(q\bar q\rightarrow t\bar t h+g)
& \stackrel{soft}{\longrightarrow} & d(PS_3)(q\bar q\rightarrow t\bar t h)
d(PS_g)_{soft}\\
&=& d(PS_3)(q\bar q\rightarrow t\bar t h)
\frac{d^{(d-1)}k}{(2\pi)^{(d-1)}2E_g} \theta(\delta_s {\sqrt{s}\over 2}-E_g)
\nonumber\,\,\,,
\end{eqnarray}
where $d(PS_g)_{soft}$ denotes the integration over the phase space of
the soft gluon. The parton level soft cross section can then be written
as:
\begin{equation}
\label{eq:sigma_soft}
\hat{\sigma}_{soft}=(4 \pi\alpha_s) \, \mu^{2 \epsilon} \,
\int d(PS_3) \overline{\sum}|{\cal A}_{\sss LO}|^2\int d(PS_g)_{soft} 
\Phi_{eik}\,\,\,.
\end{equation}
Since the contribution of the soft gluon is now completely factorized,
we can perform the integration over $d(PS_g)_{soft}$ in
Eq.~(\ref{eq:sigma_soft}) analytically, and extract the soft poles
that will have to cancel $X_{-2}^{virt}$ and $X_{-1}^{virt}$ of
Eq.~(\ref{eq:sigma_ir_poles}).  The integration over the gluon phase
space in Eq.~(\ref{eq:sigma_soft}) can be performed using standard
techniques and we refer to Refs.~\cite{Harris:2001sx,Beenakker:1989bq}
for more details.  For sake of completeness, in
Appendix~\ref{sec:soft_int} we give explicit results for the soft
integrals used in our calculation.

Finally, the soft gluon contribution to 
$\hat{\sigma}_{real}^{q\bar q}$ can be written as follows:
\begin{equation}
\label{eq:sigma_soft_total}
\hat{\sigma}_{soft}=
\frac{\alpha_s}{2\pi}{\cal N}_t
\int d(PS_3) \overline{\sum} |{\cal A}_{\sss LO}|^2 
\left\{
\frac{X_{-2}^{s}}{\epsilon^2}+\frac{X_{-1}^{s}}{\epsilon}+
N C_1^s+\frac{C_2^s}{N}\right\} \,\,\,,
\end{equation}
where
\begin{eqnarray}
\label{eq:sigma_soft_poles}
X_{-2}^{s} &=& -X_{-2}^{virt}\,\,\,,\\
X_{-1}^{s} &=& -X_{-1}^{virt}-\left(N-\frac{1}{N}\right)\,\left[\frac{3}{2} 
+2 \ln\left(\delta_s\right)\right]\,\,\,,\nonumber\\
C_1^s&=&
\frac{3}{2}\ln\left(\frac{s}{\mu^2}\right) + 2\ln^2(\delta_s)
-2\ln(\delta_s)\left[1+\ln\left(
{m_t^2\mu^2\over s_{qt} s_{\bar q\bar t}}\right)\right]
\nonumber \\
&&
+{1\over 2} \ln^2\biggl({s\over m_t^2}\biggr) -{\pi^2\over 3}
-\ln\biggl({s\over m_t^2}\biggr) \biggl[{5\over 2} 
+\ln\biggl({sm_t^2\over s_{qt} s_{\bar q\bar t}}\biggr)\biggr]
\nonumber \\
&&+{\Lambda_{t\bar t}\over \beta_{t\bar t}}
+{1\over 2} \biggl[ F_{qt}
+F_{\bar q\bar t}\biggr]+\left[ \frac{3}{2} 
+2 \ln\left(\delta_s\right)\right]\,\ln\left(\frac{\mu^2}{m_t^2}\right)
\,\,\,,\nonumber \\
C_2^s&=&
-{3\over 2} \ln \biggl({ s\over \mu^2}\biggr)-2
\ln^2(\delta_s) -2
\ln( \delta_s)\biggl[ -1 +
\frac{s_{t\bar t}}{(2 m_t^2+s_{t\bar t})\beta_{t\bar t}}
\Lambda_{t\bar t}
\nonumber \\
&&
+\ln\biggl({s\over \mu^2}\biggr)+2
\ln\biggl( {s_{qt}s_{\bar q\bar t}\over s_{q\bar t} s_{\bar qt}}\biggr)\biggr]
\nonumber \\
&&
-{1\over 2} \ln^2\biggl({s\over m_t^2}\biggr)
+{\pi^2\over 3} -\ln\biggl({s\over m_t^2}\biggr)\biggl[
-{5\over 2} +
\frac{s_{t\bar t}}{(2 m_t^2+s_{t\bar t})\beta_{t\bar t}}
\Lambda_{t\bar t}
\nonumber \\
&&
+2\ln\biggl({s_{qt} s_{\bar q\bar t}\over s_{q\bar t}
s_{\bar qt}}\biggr)
\biggr]-
{\Lambda_{t\bar t}\over \beta_{t\bar t}}
\nonumber \\
&&
+\frac{2s_{t\bar t}}{(2 m_t^2+s_{t\bar t})\beta_{t\bar t}}
\biggl[
\mbox{Li}_2\biggl({2 \beta_{t\bar t}\over 1+\beta_{t\bar t}}\biggr)
+{1\over 4} \ln^2\biggl( {1+\beta_{t\bar t}\over 
1-\beta_{t\bar t}}\biggr)\biggr]
\nonumber \\
&&
-F_{qt}+F_{q\bar t}+F_{\bar qt}
v-F_{\bar q\bar t}-\left[\frac{3}{2} 
+2 \ln\left(\delta_s\right)\right]\,\ln\left(\frac{\mu^2}{m_t^2}\right)
\,\,\, ,\nonumber
\end{eqnarray}
while ${\cal N}_t$ is defined in Eq.~(\ref{eq:nsnt}), and
$\mbox{Li}_2$ denotes the dilogarithm as described in
Ref.~\cite{lewin}.  $\beta_{t \bar t}$ and $\Lambda_{t \bar t}$ are
defined in Eq.~(\ref{eq:betadef}), while, for any initial parton $i$
and final parton $f$, the function $F_{if}$ can be written as:
\begin{eqnarray}
\label{eq:f_if}
F_{if}&=&
\ln^2\biggl({1-\beta_f\over 1-\beta_f \cos \theta_{if}}\biggr)
-{1\over 2} \ln^2\biggl( {1+\beta_f\over 1-\beta_f}\biggr)\\
&&+2 \mbox{Li}_2\biggl(-{\beta_f (1-\cos \theta_{if})\over 
1-\beta_f}\biggr)
-2 \mbox{Li}_2\biggl(-{\beta_f
 (1+\cos \theta_{if})\over 1-\beta_f \cos\theta_{if}}\biggr)\,\,\,,
\nonumber
\end{eqnarray}
where $\cos \theta_{if}$ is the angle between partons $i$ and $f$ in
the center-of-mass frame of the initial state partons, and
\begin{equation}
\beta_f=\sqrt{1-\frac{m_t^2}{(p_f^0)^2}}\,\,\,\,\,,\,\,\,\,\, 
1-\beta_f\cos{\theta_{if}}=\frac{s_{if}}{p_f^0\sqrt{s}}\,\,\,.
\end{equation}
All the quantities in
Eq.~(\ref{eq:f_if}) can be expressed in terms of kinematic Al
invariants, once we use $s_{if}\!=\!2p_i\!\cdot\!p_f$ together with:
\begin{equation}
p_t^0=\frac{s-{\bar s}_{{\bar t}h}+m_t^2}{2\sqrt{s}}
\,\,\,\,\,\mbox{and}\,\,\,\,\,
p_{\bar t}^0=\frac{s-{\bar s}_{th}+m_t^2}{2\sqrt{s}}\,\,\,,
\end{equation}
where ${\bar s}_{fh}\!=\!(p_f+p_h)^2$.  As can be easily seen from
Eqs.~(\ref{eq:sigma_ir_poles}) and (\ref{eq:sigma_soft_poles}), the IR
poles of the virtual corrections are exactly canceled by the
corresponding singularities in the soft gluon contribution. The
remaining IR poles in $\hat \sigma_{soft}$ will be canceled by the PDF
counterterms as described in detail in Sec.~\ref{sec:total}.

\subsubsection{Hard gluon emission} 
\label{subsubsec:two_cutoff_hard}

The hard region of the gluon phase space is defined by requiring 
that the energy of the emitted gluon is above a given threshold.
As we discussed earlier this is expressed by the condition that
\begin{equation}
E_g >\delta_s {\sqrt{s}\over 2}\,\,\,,
\end{equation}
for an arbitrary small \emph{soft} cutoff $\delta_s$, which
automatically assures that $\hat{\sigma}_{hard}$ does not contain soft
singularities.  However, a hard gluon can still give origin to
singularities when it is emitted at a small angle, i.e.
\emph{collinear}, to a massless incoming or outgoing parton. In order
to isolate these divergences and compute them analytically, we further
divide the hard region of the $q\bar q\rightarrow t\bar th+g$ phase
space into a \emph{hard/collinear} and a \emph{hard/non-collinear}
region, by introducing a second small \emph{collinear} cutoff
$\delta_c$.  The \emph{hard/non-collinear} region is defined by the
condition that both
\begin{equation}
\label{eq:deltac_cuts}
\frac{2 q_1\!\cdot\! k}{E_g \sqrt{s}}> \delta_c\,\,\,\,\,\,\,
\mbox{and}\,\,\,\,\,\,\,
\frac{2q_2\!\cdot\! k}{E_g\sqrt{s}}>\delta_c
\end{equation}
are verified. The contribution from the \emph{hard/non-collinear}
region, $\hat{\sigma}_{hard/non-coll}$, is finite and we compute it
numerically by using standard Monte Carlo integration techniques.

In the $\emph{hard/collinear}$ region, one of the conditions in
Eq.~(\ref{eq:deltac_cuts}) is not satisfied and the hard gluon is
emitted collinear to one of the incoming partons.  In this region, the
initial-state parton $i$ ($i\!=\!q,\bar q$) is considered to split
into a hard parton $i^\prime$ and a collinear gluon $g$, $i\rightarrow
i^\prime g$, with $p_{i^\prime}\!=\!z p_i$ and $k\!=\!(1-z)p_i$. The
matrix element squared for $i j\rightarrow t\bar t h+g$ factorizes
into the Born matrix element squared and the Altarelli-Parisi
splitting function for $i\rightarrow i^\prime g$, i.e.:
\begin{equation}
\label{eq:m2_coll_lim}
\overline{\sum}|{\cal A}_{real}(ij \rightarrow t\bar t h+g)|^2 
\stackrel{collinear}{\longrightarrow}
(4 \pi \alpha_s) \sum_i\overline{\sum}|{\cal A}_{\sss LO}
(i^\prime j\rightarrow t \bar th)|^2
\frac{2 P_{ii^\prime}(z)}{z \, s_{ig}} \,\,\,,
\end{equation}
with $s_{ig}=2 p_i\!\cdot\! k$.
In our case:
\begin{equation}
\label{eq:p_qq}
P_{ii^\prime}(z)=P_{qq}(z)=
C_F \left(\frac{1+z^2}{1-z}-\epsilon(1-z)\right)
\end{equation}
is the unregulated Altarelli-Parisi splitting function for
$q\!\rightarrow\!q+g$ at lowest order, including terms of ${\cal
O}(\epsilon)$, and $C_F\!=\!(N^2-1)/2N$.  Moreover, in the collinear
limit, the $q\bar q\rightarrow t\bar th+g$ phase space also factorizes as:
\begin{eqnarray}
\label{eq:ps_coll_lim}
d(PS_4)(ij\rightarrow t\bar t h+g)
&\stackrel{collinear}{\longrightarrow}& 
d(PS_3)(i^\prime j\rightarrow t\bar t h)
\frac{z\,d^{(d-1)}k}{(2\pi)^{(d-1)}2E_g}
\theta\left(E_g-\delta_s {\sqrt{s}\over 2}\right)\times \\
&&\theta(\cos\theta_{ig}-(1-\delta_c))\nonumber\\
&&\!\!\!\!\!\!\!\!\!\!\!\!\!\!\!\!\!\!\!\!\!\! \stackrel{d=4-2 \epsilon}{=}
\frac{\Gamma(1-\epsilon)}{\Gamma(1-2\epsilon)}
\frac{\left(4\pi\right)^\epsilon}{16 \pi^2}\,z\,dz\,ds_{ig} 
\left[(1-z) s_{ig}\right]^{-\epsilon}
\theta\left({(1-z)\over z }s^\prime {\delta_c \over 2}-s_{ig}\right)
\nonumber\; ,
\end{eqnarray}
where the integration range for $s_{ig}$ in the collinear region is
given in terms of the collinear cutoff, and we have defined
$s^\prime\!=\!2 p_{i^\prime}\cdot p_j$.  The integral over the
collinear gluon degrees of freedom can then be performed separately,
and this allows us to explicitly extract the collinear singularities
of $\hat{\sigma}_{hard}$.  $\hat{\sigma}_{hard/coll}$ turns out to be
of the form \cite{Harris:2001sx,Baur:1999kt}:
\begin{eqnarray}
\label{eq:coll_pole}
\hat{\sigma}_{hard/coll}&=&
\left[\frac{\alpha_s}{2\pi}\frac{\Gamma(1-\epsilon)}{\Gamma(1-2\epsilon)}
\left(\frac{4\pi\mu^2}{m_t^2}\right)^\epsilon\right]
\left(-\frac{1}{\epsilon}\right)\delta_c^{-\epsilon}\times\\
&&\left\{
\int_{0}^{1-\delta_s} dz
\left[\frac{(1-z)^2}{2 z} \frac{s^\prime}{m_t^2}\right]^{-\epsilon} 
P_{ii^\prime}(z) \,
\hat{\sigma}_{\sss LO}(i^\prime j\rightarrow t\bar t h)
+ (i\leftrightarrow j)\right\}\,\,\,.\nonumber
\end{eqnarray}
The upper limit on the $z$ integration ensures the exclusion of the
soft gluon region.  As usual, these initial-state collinear
divergences are absorbed into the parton distribution functions as
will be described in detail in the Sec.~\ref{sec:total}.

\subsection{Phase Space Slicing method with one cutoff}
\label{subsec:one_cutoff}

An alternative way of isolating both soft and collinear singularities
is to divide the phase space of the final state partons into two
regions according to whether all partons can be resolved (the
\emph{hard} region) or not (the \emph{infrared}, or \emph{ir},
region). In the case of $q\bar q\rightarrow t\bar th+g$, the
\emph{hard} and \emph{ir} regions are defined by whether the gluon is
resolved or not.  The emitted gluon is not resolved, and therefore
considered \emph{ir}, when
\begin{equation}
\label{eq:smin_condition}
s_{ig}=2p_i\cdot k<s_{min}\,\,\,,\,\,\,\mbox{with}
\,\,\,\,\,i=q,\bar q,t,\bar t\,\,\,,
\end{equation}
for an arbitrary small cutoff $s_{min}$. Similarly to
Eq.~(\ref{eq:sigma_real_two_cutoff}), the partonic real cross section
can be written as the sum of two terms:
\begin{equation}
\label{eq:sigma_real_one_cutoff}
\hat\sigma_{real}^{q\bar q}=\hat\sigma_{ir}+\hat\sigma_{hard}\,\,\,,
\end{equation}
where $\hat\sigma_{ir}$ includes both soft and collinear
singularities, while $\hat\sigma_{hard}$ is finite. Following the
general idea of PSS, we calculate $\hat\sigma_{ir}$ analytically,
while we evaluate $\hat\sigma_{hard}$ numerically, using standard
Monte Carlo integration techniques. Both $\hat\sigma_{ir}$ and
$\hat\sigma_{hard}$ depend on the cutoff $s_{min}$, but the hadronic
real cross section, $\sigma_{real}$, is cutoff independent, after mass
factorization, as will be shown in Sec.~\ref{sec:total}.

In order to calculate $\hat\sigma_{ir}$ we apply the formalism
developed in Refs.~\cite{Giele:1992vf,Giele:1993dj,Keller:1999tf}
as follows.
\begin{itemize}
\item We consider the crossed process $h\rightarrow q\bar qt\bar t+g$
  which is obtained from $q\bar q\rightarrow t\bar th+g$ by crossing
  all the initial state colored partons to the final state, while
  crossing the Higgs boson to the initial state.  For a systematic
  extraction of the IR singularities within the one-cutoff method, we
  organize the amplitude for $h\rightarrow q\bar qt\bar t+g$, ${\cal
    A}^{h\to q\bar qt\bar tg}$, in terms of colored ordered amplitudes
  \cite{Berends:1989yn}.  Using the color decomposition:
\begin{equation}
\label{eq:tata}
T^a_{c_1c_2}T^a_{c_3c_4}=\frac{1}{2}\left(\delta_{c_1c_4}\delta_{c_3c_2}-
\frac{1}{N}\delta_{c_1c_2}\delta_{c_3c_4}\right)\,\,\,,
\end{equation}
we write ${\cal A}^{h\to q\bar qt\bar tg}$ as the sum of four
color ordered amplitudes ${\cal A}_1,\ldots,{\cal A}_4$ as follows:
\begin{eqnarray}
\label{eq:a_real_a1a2a3a4}
{\cal A}^{h\to q\bar qt\bar tg}&=&
ig_s\delta_{f_qf_{\bar q}}\delta_{f_tf_{\bar t}}
\frac{1}{2}\left(
\delta_{c_tc_{\bar q}}T^a_{c_qc_{\bar t}}{\cal A}_1(p_t,p_t^\prime,q_1,q_2,k)+
T^a_{c_tc_{\bar q}}\delta_{c_qc_{\bar t}}{\cal A}_2(p_t,p_t^\prime,q_1,q_2,k)
\right.\nonumber\\
&-&\left.\frac{1}{N}
\delta_{c_tc_{\bar t}}T^a_{c_qc_{\bar q}}{\cal A}_3(p_t,p_t^\prime,q_1,q_2,k)
-\frac{1}{N}
T^a_{c_tc_{\bar t}}\delta_{c_qc_{\bar q}}{\cal A}_4(p_t,p_t^\prime,q_1,q_2,k)
\right)\,\,\,,
\end{eqnarray}
where $g_s\!=\!\sqrt{4\pi\alpha_s}$, while $(f_q,f_{\bar
  q},f_t,f_{\bar t})$ and $(c_q,c_{\bar q},c_t,c_{\bar t})$ denote the
flavor and color indices of the various outgoing quarks.  The
amplitudes ${\cal A}_i(p_t,p_t^\prime,q_1,q_2,k)$ (for $i=1,2,3,4$)
correspond to the four possible independent color structures that
arise in the $h\rightarrow t\bar t q\bar q+g$ process, and each ${\cal
  A}_i$ contains terms describing the emission of the gluon from a
different pair of external quarks. We give the explicit expressions
for the ${\cal A}_i$ amplitudes in Appendix \ref{sec:a_real_a1a2a3a4}.
Due to this decomposition, the partonic cross section for $h\to q\bar
q t \bar t+g$ can be written in a very compact form:
\begin{equation}
\label{eq:sigreal_crossed}
\hat\sigma^{h\to q\bar qt\bar tg}=\int d(PS_5) 
\overline{\sum}|{\cal A}^{h\to q\bar qt\bar tg}|^2\,\,\,, 
\end{equation}
with
\begin{eqnarray}
\label{eq:a2_real_a1a2a3a4}
\overline{\sum}|{\cal A}^{h\to q\bar qt\bar tg}|^2 
&=& \left({g_s^2N\over 2}\right)  
\left(\frac{N^2-1}{4}\right)\overline{\sum}
\left\{|{\cal A}_1|^2+|{\cal A}_2|^2+\right.\\
&& \left.\frac{1}{N^2}\left[-2|{\cal A}_3+{\cal A}_4|^2+
|{\cal A}_3|^2+|{\cal A}_4|^2\right]\right\}\,\,\,.\nonumber
\end{eqnarray}
\item Using the one-cutoff PSS method and the factorization properties
  of both the color ordered amplitudes ${\cal A}_i$ and the gluon
  phase space in the soft/collinear limit, we extract the IR
  singularities of $\hat\sigma^{h\to q\bar qt\bar tg}$ into $\hat
  \sigma_{soft}^{h\to q\bar qt\bar tg}$ and $\hat \sigma_{coll}^{h\to
    q\bar qt\bar tg}$ as follows:
\begin{eqnarray}
\label{eq:sigsoft_crossed}
\hat \sigma^{h\to q\bar qt\bar tg}& \stackrel{soft}{\longrightarrow}& 
\hat \sigma_{soft}^{h\to q\bar qt\bar tg}
=\int d(PS_4) d(PS_g)_{soft} \overline{\sum}
|{\cal A}_{soft}^{h\to q\bar qt\bar tg}|^2\,\,\,,
\\
\label{eq:sigsoll_crossed}
\hat \sigma^{h\to q\bar qt\bar tg}& \stackrel{collinear}{\longrightarrow}& 
\hat \sigma_{coll}^{h\to q\bar qt\bar tg}
=\int d(PS_4) d(PS_g)_{coll} \overline{\sum}
|{\cal A}_{coll}^{h\to q\bar qt\bar tg}|^2
\; ,
\end{eqnarray}
where we denote by $d(PS_g)_{soft}$ ($d(PS_g)_{coll}$) the phase space
of the gluon in the soft (collinear) limit, while
$\overline{\sum}|{\cal A}_{soft}^{h\to q\bar qt\bar tg}|^2$
($\overline{\sum}|{\cal A}_{coll}^{h\to q\bar qt\bar tg}|^2$)
represents the soft (collinear) limit of
Eq.(\ref{eq:a2_real_a1a2a3a4}). The explicit calculation of
$\hat\sigma_{soft,coll}^{h\to q\bar qt\bar tg}$ is described in detail
in Sections~\ref{subsubsec:one_cutoff_soft} and
\ref{subsubsec:one_cutoff_collinear}, respectively.  The factorization
of soft and collinear singularities for color ordered amplitudes has
been discussed in the literature mainly for the leading color terms
(${\cal O}(N)$).  For our application of the one-cutoff PSS method, we
will have to extend these results to the sub-leading color terms
(${\cal O}(1/N)$).
\item
Finally, the IR singular contribution $\hat \sigma_{ir}$ in
Eq.~(\ref{eq:sigma_real_one_cutoff}) consists of two terms:
\begin{equation}
\hat \sigma_{ir}=\hat \sigma_{ir}^c+\hat \sigma_{crossing} \; .
\end{equation}
As described in detail in Sec.~\ref{subsubsec:one_cutoff_ir} , $\hat
\sigma^c_{ir}$ is obtained by crossing $q$ and $\bar q$ to the initial
state and $h$ to the final state in the sum of $\hat
\sigma_{soft}^{h\to q\bar qt\bar tg}$ and $\hat \sigma_{coll}^{h\to
  q\bar qt\bar tg}$, while $\hat \sigma_{crossing}$ corrects for the
difference between the collinear gluon radiation from initial and
final state partons \cite{Giele:1993dj}, as will be discussed in
detail in Sec.~\ref{sec:total}.  As explicitly shown in
Sec.~\ref{subsubsec:one_cutoff_ir}, the IR singularities of $\hat
\sigma_{virt}^{q\bar q}$ of Sec.~\ref{subsec:virtual_ir} are exactly
canceled by the corresponding singularities in $\hat\sigma_{ir}^c$. On
the other hand, $\hat \sigma_{crossing}$ still contains collinear
divergences that will be canceled by the PDF counterterms when the
parton cross section is convoluted with the PDFs (see
Sec.~\ref{sec:total}).
\end{itemize}

\subsubsection{Soft gluon emission}
\label{subsubsec:one_cutoff_soft}

We first consider the case of soft singularities, when, in the limit
of $E_g\!\rightarrow\!0$ (soft limit), one or more
$s_{ig}\!<\!s_{min}$ ($i\!=\!q,\bar q,t,\bar t$). Using the
factorization properties of the color ordered amplitudes ${\cal A}_i$
in the soft limit, the amplitude squared for $h \to q\bar qt\bar t+g$
can be written as:
\begin{eqnarray}
\label{eq:a2_soft_fab}
\overline{\sum}|{\cal A}^{h\to q\bar qt\bar tg}|^2
&\stackrel{soft}{\longrightarrow}& 
\overline{\sum}|{\cal A}^{h\to q\bar qt\bar tg}_{soft}|^2=
\left(\frac{g_s^2 N}{2}\right)
\overline{\sum}|{\cal A}_{\sss LO}^{h \to q\bar qt\bar t}|^2   
\left\{\phantom{\frac{1}{N}}\!\!\!
f_{q\bar t}(g)+f_{\bar qt}(g)- \right.\\
&&\left.\frac{1}{N^2}\left[f_{t\bar t}(g)+f_{q\bar q}(g)
-2\left(f_{qt}(g)-f_{q\bar t}(g)-f_{\bar q t}(g)+
f_{\bar q\bar t}(g)\right)\right]\right\}\,\,\,,\nonumber
\end{eqnarray}
where, for any pair of quarks $(a,b)$, the soft functions
$f_{ab}(g)$ are defined as:
\begin{equation}
\label{eq:fab}
f_{ab}(g)\equiv {4 s_{ab}\over s_{ag}s_{bg}} -{4m_a^2\over s_{ag}^2}
-{4 m_b^2\over s_{bg}^2}\,\,\,,
\end{equation}
and, as before (see Eq.~(\ref{eq:ir_invariants})),
\[
s_{ij}\equiv 2 p_i\cdot p_j\,\,\,,
\]
both for massless and massive quarks. ${\cal A}_{\sss LO}^{h \to q\bar
  qt\bar t}$ is the tree level amplitude for the process $h \to q\bar
qt\bar t$ as given by Eq.~(\ref{eq:a_lo_a0}).  We note that
Eq.~(\ref{eq:a2_soft_fab}) corresponds to the factorization property
expressed in Eq.~(\ref{eq:m2_soft_lim}). Since, in the soft limit,
the $h\to q\bar q t\bar t+g$ phase space also factorizes, in analogy
to Eq.~(\ref{eq:ps_soft_lim}), we can integrate out the soft gluon
degrees of freedom and obtain the soft gluon part of the cross section
for $h\to q\bar qt\bar t+g$ as:
\begin{equation}
\label{eq:a2_soft}
\hat \sigma^{h\to q\bar qt\bar tg}_{soft}=\int d(PS_4)
\overline{\sum}|{\cal A}_{\sss LO}^{h\to q\bar qt\bar t}|^2
\left\{ S_{q\bar t}+S_{\bar qt}-
\frac{1}{N^2}\left[S_{t\bar t}+S_{q\bar q}
-2\left(S_{qt}-S_{q\bar t}-S_{\bar qt}+S_{\bar q\bar t}\right)\right]
\right\}\,\,\,,
\nonumber
\end{equation}
where, for any pair of quarks $(a,b)$, the integrated soft functions
$S_{ab}$ are defined as:
\begin{equation}
\label{eq:sab}
S_{ab}= \frac{g_s^2N}{2}\int d(PS_g)_{soft}(a,b,g) f_{ab}(g) \,\,\,.
\end{equation}
In the one-cutoff PSS method, the explicit form of the soft gluon phase
space integral is given by \cite{Keller:1999tf}:
\begin{eqnarray}
\label{eq:ps_soft}
d(PS_g)_{soft}(a,b,g)&=&\frac{(4\pi)^\epsilon}{16\pi^2}
\frac{\lambda^{\left(\epsilon-\frac{1}{2}\right)}}{\Gamma(1-\epsilon)}
\left[s_{ag}s_{bg}s_{ab}-m_b^2s_{ag}^2-m_a^2s_{bg}^2\right]^{-\epsilon}
ds_{ag}ds_{bg}\times\nonumber\\
&&\theta(s_{min}-s_{ag})\theta(s_{min}-s_{bg})\,\,\,,
\end{eqnarray}
where
\begin{equation}
\lambda=s_{ab}^2-4m_a^2m_b^2 \; ,
\end{equation}
and the integration boundaries for $s_{ag}$ and $s_{bg}$ vary
accordingly to whether $a$ and $b$ are massive or massless
quarks (see Ref.~\cite{Keller:1999tf} for more details).

The explicit form of the integrated soft functions $S_{ab}$ is
obtained by carrying out the integration in Eq.~(\ref{eq:sab}).
When $a\!=\!q$ and $b\!=\!\bar q$, i.e. when both quarks are massless,
the integrated soft function $S_{q\bar q}$ is given by \cite{Giele:1992vf}:
\begin{equation}
\label{eq:soft_function_qqbar}
S_{q\bar q}=
\left(\frac{\alpha_s N}{2\pi}\right)\frac{1}{\Gamma(1-\epsilon)}
\left(\frac{4\pi\mu^2}{s_{min}}\right)^\epsilon
\frac{1}{\epsilon^2}\left(\frac{s}{s_{min}}\right)^\epsilon\,\,\,,
\end{equation}
where, in our notation, $s\!=\!s_{q\bar q}$ is the parton
center-of-mass energy (see Eq.~(\ref{eq:ir_invariants})). On the other
hand, when $a\!=\!q,\bar q$ and $b\!=\!t,\bar t$, i.e. when one quark
is massless and the other is massive, the corresponding integrated
soft functions are of the form \cite{Keller:1999tf}:
\begin{eqnarray}
S_{ab}&=&
\label{eq:soft_function_ab}
\left(\frac{\alpha_sN}{2\pi}\right)\frac{1}{\Gamma(1-\epsilon)}
\left(\frac{4\pi\mu^2}{s_{min}}\right)^\epsilon
\biggl({s_{ab}\over s_{min}}\biggr)^\epsilon\times\nonumber \\
&&
\biggl\{
\frac{1}{\epsilon^2}\biggl(1-{1\over 2} 
\biggl(\frac{s_{ab}}{m_t^2}\biggr)^\epsilon\biggr)
+{1\over 2 \epsilon}\biggl({s_{ab}\over m_t^2}\biggr)^\epsilon
-{1\over 2} \zeta(2)+{m_t^2\over s_{ab}}\biggr\}
\nonumber \\ 
&=&
{\alpha_s N\over 2 \pi} {1\over \Gamma(1-\epsilon)}
\left(\frac{4\pi\mu^2}{s_{min}}\right)^\epsilon
\biggl\{ {1\over 2 \epsilon^2} +{1\over 2 \epsilon} +{1\over 2 \epsilon}
\ln\biggl({m_t^2\over s_{min}}\biggr)\\
&& +{1\over 4}  \ln^2\biggl({m_t^2\over s_{min}}\biggr)
-{1\over 2}\ln^2\biggl({s_{ab}\over m_t^2}\biggr)
+{1\over 2}\ln\biggl({s_{ab}\over m_t^2}\biggr)
\nonumber \\ &&
+{1\over 2}\ln\biggl({s_{ab}\over s_{min}}\biggr)
-{1\over 2} \zeta(2)+{m_t^2\over s_{ab}}\biggr\}\,\,\,.\nonumber
\end{eqnarray}
Finally, when $a\!=\!t$ and $b\!=\!\bar t$, i.e. when both quarks are
massive, the corresponding integrated soft function $S_{t\bar t}$ is
given by \cite{Keller:1999tf}:
\begin{equation}
\label{eq:soft_function_ttbar}
S_{t\bar t}=\left(\frac{\alpha_sN}{2 \pi}\right)
\frac{1}{\Gamma(1-\epsilon)}
\left(\frac{4\pi\mu^2}{s_{min}}\right)^\epsilon 
\frac{m_t^2}{\sqrt{\lambda_{t\bar t}}}
\left(J_{s}\frac{1}{\epsilon}+J_a+J_b\right)\,\,\,,
\end{equation}
where we have defined:
\begin{eqnarray}
\label{eq:ttbar_jsjajb}
{m_t^2\over \sqrt{\lambda_{t\bar t}}}J_{s}&=&
1-{s_{t\bar t}\over (2 m_t^2+s_{t\bar t})\beta_{t\bar t}}
\Lambda_{t\bar t}\,\,\,,
\nonumber \\
J_a&=& J_{s} 
\ln\biggl({\tau_+^2\lambda_{t\bar t}\over s_{min} m_t^2}\biggr)
\,\,\,,\nonumber \\
J_b&=&\biggl(\tau_+-\tau_-\biggr)
\biggl[1-2\ln(\tau_+-\tau_-)-\ln(\tau_+)\biggr]\nonumber \\
&&+\biggl({\tau_++\tau_-\over 2}\biggr)
\biggl[\ln({\tau_+\over \tau_-})
\biggl(1+2\ln(\tau_+-\tau_-)\biggr) \\
&&+\mbox{Li}_2\biggl(1-{
\tau_+\over \tau_-}\biggr)-\mbox{Li}_2\biggl(1-{
\tau_-\over \tau_+}\biggr)\biggr]
\nonumber \\
&&
+1+\tau_-\tau_++(\tau_-+\tau_+)\biggl[
-1-\ln(\tau_+)\ln(\tau_-)+{1\over 2}\ln^2(\tau_+)
\biggr]\,\,\, .\nonumber
\end{eqnarray}
$\beta_{t\bar t}$ and $\Lambda_{t\bar t}$ are defined in
Eq.~(\ref{eq:betadef}) while $\lambda_{t\bar t}$ and $\tau_{\pm}$ are
given by:
\begin{eqnarray}
\lambda_{t\bar t} &\equiv& s_{t\bar t}^2-4 m_t^4\,\,\,,\nonumber \\
\tau_{\pm}&=&\frac{s_{t\bar t}}{2m_t^2}\pm
\sqrt{\left(\frac{s_{t\bar t}}{2m_t^2}\right)^2-1}
\nonumber\,\,\,.
\end{eqnarray}
Finally, using
Eqs.~(\ref{eq:soft_function_qqbar})-(\ref{eq:ttbar_jsjajb}), we can
derive the complete form of $\hat \sigma^{h\to q\bar qt\bar
  tg}_{soft}$:
\begin{eqnarray}
\label{eq:a2_soft_tot}
\hat \sigma^{h\to q\bar qt\bar tg}_{soft}&=&
{\alpha_s \over 2 \pi} {1\over \Gamma(1-\epsilon)}
\left(\frac{4\pi\mu^2}{m_t^2}\right)^\epsilon
\int d(PS_4) \overline{\sum}|{\cal A}^{h\to q\bar qt\bar t}_{\sss LO}|^2
\biggr\{ {\tilde X_{-2}^s\over \epsilon^2}+{\tilde X_{-1}^s\over \epsilon}
+N \tilde C^s_1+\frac{\tilde C_2^s}{N}
\biggr\}\,\,\,,
\nonumber\\
\end{eqnarray}
where
\begin{eqnarray}
\tilde X_{-2}^s &=&\biggl(N-{1\over N}\biggr)\,\,\,,\nonumber\\
\tilde X_{-1}^s &=&N\biggl[1+
2\ln \biggl({m_t^2\over s_{min}}\biggr)\biggr]
-{1\over N}\left[
\ln\biggl({s \over s_{min}}\biggr)
+\ln\left(\frac{m_t^2}{s_{min}}\right)+
1-\frac{s_{t\bar t}}
{(2m_t^2+s_{t\bar t})\beta_{t\bar t}} \Lambda_{t\bar t}\right] \; ,
\nonumber \\
\tilde C^s_1&=&2 \ln^2\biggl({m_t^2\over s_{min}}\biggr)+
\ln\left(\frac{m_t^2}{s_{min}}\right)
-{1\over 2}\ln^2\biggl({s_{q\bar t}\over m_t^2}\biggr) 
-{1\over 2}\ln^2\biggl({s_{\bar qt}\over m_t^2}\biggr)
+\ln\biggl({s_{q\bar t}s_{\bar qt}\over m_t^2 s_{min}}\biggr)
\nonumber \\
&&
-\zeta(2)+m_t^2\biggl({1\over s_{q\bar t}}+{1\over s_{\bar qt}}\biggr)
\,\,\,,\\
\tilde C^s_2&=&-\left[
\frac{1}{2}\ln^2\left(\frac{m_t^2}{s_{min}}\right)+
\ln\left(\frac{m_t^2}{s_{min}}\right)\left(
\ln\left(\frac{s}{s_{min}}\right)+
1-\frac{s_{t\bar t}}{(2m_t^2+s_{t\bar t})\beta_{t\bar t}}
\Lambda_{t\bar t} \right)\right.\nonumber\\
&&+\frac{1}{2}\ln^2\biggl({s\over s_{min}}\biggr)+
{m_t^2\over \sqrt{\lambda_{t\bar t}}}
\left(J_a+J_b\right)+
\ln ^2\biggl({s_{qt}\over m_t^2}\biggr)
-\ln ^2\biggl({s_{q\bar t}\over m_t^2}\biggr) 
-\ln ^2\biggl({s_{\bar qt}\over m_t^2}\biggr) 
\nonumber \\
&&\left.
+\ln ^2\biggl({s_{\bar q\bar t}\over m_t^2}\biggr)
-2\ln\biggl({s_{qt}s_{\bar q\bar t}\over s_{\bar qt}s_{q\bar t}}\biggr)
-2m_t^2\biggl({1\over s_{qt}}-{1\over s_{q\bar t}}-{1\over s_{\bar qt}}
+{1\over s_{\bar q\bar t}}\biggr)\right]\,\,\,.\nonumber
\end{eqnarray}

\subsubsection{Collinear gluon emission}
\label{subsubsec:one_cutoff_collinear}

In the collinear limit when an external massless quark ($i$) and a
hard gluon become collinear and cluster to form a new parton
($i^\prime$) ($i+g\rightarrow i^\prime $, with collinear kinematics:
$p_i\!=\!zp_{i^\prime}$ and $k\!=\!(1-z)p_{i^\prime}$), the color
ordered amplitudes factorize and the amplitude squared for
$h\rightarrow q\bar qt\bar t+g$ can be written as:
\begin{eqnarray}
\label{eq:a2_collinear_fij}
\overline{\sum}|{\cal A}^{h\to q\bar qt\bar tg}|^2
&\stackrel{collinear}{\longrightarrow}& 
\overline{\sum}|{\cal A}^{h\to q\bar qt\bar tg}_{coll}|^2=
\left(\frac{g_s^2 N}{2}\right)
\overline{\sum}|{\cal A}_{\sss LO}^{h \to q\bar qt\bar t}|^2   
\left\{\phantom{\frac{1}{N}}\!\!\!
f^{qg\to q}_{\bar t}+f^{\bar qg\to\bar q}_{t} \right.\\
&-&\left.\frac{1}{N^2}\left[f^{qg\to q}_{\bar q}+
f^{\bar qg\to\bar q}_{q}
-2\left(f^{qg\to q}_{t}-f^{qg\to q}_{\bar t}-
f^{\bar q\to\bar q}_{t}+f^{\bar qg\to\bar q}_{\bar t}\right)\right]
\right\}\,\,\,.\nonumber
\end{eqnarray}
The collinear functions $f^{ig\to i^\prime}_{j}$ contain
the collinear singularity and are proportional to the Altarelli-Parisi
splitting function for $ig\rightarrow i^\prime$ (see
Eq.~(\ref{eq:p_qq})), i.e.:
\begin{equation}
\label{eq:fij_c}
f^{ig\to i^\prime}_{j}\equiv \frac{2}{s_{ig}}
\left(\frac{1+z^2}{1-z}-\epsilon(1-z)\right) \,\,\,.
\end{equation}
Using this definition, we can see that Eq.~(\ref{eq:a2_collinear_fij})
is equivalent to Eq.~(\ref{eq:m2_coll_lim}), although $q$ and $\bar
q$, the massless quarks, are now considered as final state quarks.
The reason why we use a more involved expression is because this
allows us to match the collinear and soft regions of the gluon phase
space in a very natural way, as will be explained in the following. In
the same spirit, the lower index $j$ of the collinear functions
$f^{ig\to i^\prime}_{j}$ keeps track of which color ordered amplitude
a given collinear pole comes from. Although seemingly useless at this
stage, this will be crucial in deriving Eqs.~(\ref{eq:a2_coll}) and
(\ref{eq:coll_function}), where the integration over the collinear
region of the gluon phase space is performed in such a way to avoid to
overlap with the soft gluon phase space integration in
Eqs.~(\ref{eq:a2_soft}) and (\ref{eq:sab}). Finally, we note that
there is no $f^{tg\to t}_{\bar t}$ or $f^{\bar tg\to\bar t}_{t}$ in
Eq.~(\ref{eq:a2_collinear_fij}) since the gluon emission from a
massive quark does not give origin to collinear singularities.

In the collinear limit, the $h\to q\bar q t\bar t+g$ phase space also
factorizes, in complete analogy to Eq.~(\ref{eq:ps_coll_lim}),
provided the obvious changes between initial and final state partons
are taken into account.  Therefore, we can integrate out analytically
the collinear gluon degrees of freedom and obtain the collinear part
of the partonic cross section for $h\to q\bar qt\bar t+g$ as:
\begin{equation}
\label{eq:a2_coll}
\hat\sigma^{h\to q\bar qt\bar tg}_{coll}=\int d(PS_4) 
\overline{\sum}|{\cal A}^{h\to q\bar qt\bar t}_{\sss LO}|^2\left\{ 
C_{q\bar t}+C_{\bar q t}- 
\frac{1}{N^2}\left[
C_{q\bar q}-2\left(C_{qt}-C_{q\bar t}-C_{\bar qt}+C_{\bar q\bar t} 
\right)\right]\right\}\,\,\,,
\end{equation}
where, for any pair of quarks $(i,j)$, the integrated collinear
functions $C_{ij}$ are defined as:
\begin{equation}
\label{eq:coll_function}
C_{ij}=\left(\frac{g_s^2N}{2}\right)\int d(PS_g)_{coll}(i,j,z)
f^{ig\rightarrow i^\prime}_j(z)
=-\left(\frac{\alpha_sN}{2\pi}\right)
\frac{1}{\Gamma(1-\epsilon)}\left(\frac{4\pi\mu^2}{s_{min}}\right)^\epsilon
\frac{1}{\epsilon}I_{ig\rightarrow i^\prime}(z_1,z_2)\,\,\,.
\end{equation}
The phase space of the collinear gluon can be written as:
\begin{equation}
\label{eq:ps_coll}
d(PS_g)_{coll}(i,j,z)=\frac{(4\pi)^\epsilon}{16\pi^2}
\frac{1}{\Gamma(1-\epsilon)}s_{ig}^{-\epsilon}ds_{ig}
[z(1-z)]^{-\epsilon}dz \, \theta(s_{min}-s_{ig})\,\,\,,
\end{equation}
and the integration boundaries on $z$ are defined by the requirement
that only one $s_{ig}$ verifies the condition $s_{ig}\!<\!s_{min}$.
This is necessary in order to avoid overlapping with the region of
phase space where the gluon is soft (see Eq.~(\ref{eq:sab})), and it
is easily translated into an upper bound on the $z$ integration,
thanks to the structure of Eqs.~(\ref{eq:a2_soft}) and
(\ref{eq:a2_coll}).  In fact, each term in Eqs.~(\ref{eq:a2_soft}) and
(\ref{eq:a2_coll}) depends on only two invariants, $s_{ig}$ and
$s_{jg}$, and each term in $\hat\sigma_{coll}^{h\rightarrow q\bar q
  t\bar th}$ corresponds to an analogous term in $\hat
\sigma_{soft}^{h\to q\bar q t \bar tg}$ (except that $C_{t\bar t}$ is
missing since there is no collinear emission from $t$ and $\bar t$).
Therefore, for each $C_{ij}$ we only need to require that when
$s_{ig}\!<\!s_{min}$:
\begin{equation}
\label{eq:z_boundaries}
s_{jg}=(1-z)s_{ij}^\prime>s_{min} \longrightarrow z<
1-\frac{s_{min}}{s_{ij}^\prime}=1-z_2\,\,\,.
\end{equation}
The lower bound on $z$ is not constrained and the integration starts
at $z_1\!=\!0$. For sake of simplicity, and since this does not give
origin to ambiguities, in the following we will denote the
$s_{ij}^\prime$ invariants in Eq.~(\ref{eq:z_boundaries}) by $s_{ij}$.
Finally, when the integration over the collinear gluon degrees of
freedom is performed, one finds that the $I_{ig\rightarrow
  i^\prime}(z_1,z_2)$ functions in Eq.~(\ref{eq:coll_function}) are of
the form \cite{Giele:1992vf}:
\begin{equation}
 I_{ig\rightarrow i^\prime}(z_1,z_2)=
\left[\left(\frac{z_2^{-\epsilon}-1}{\epsilon}\right)-\frac{3}{4}+
  \left(\frac{\pi^2}{6}-\frac{7}{4}\right)\,\epsilon\right]+
O(\epsilon^2)\,\,\,.
\end{equation}
When $i\!=\!q,\bar q$ and $j\!=\!t,\bar t$, i.e. when one quark is
massless and the other is massive, the integrated collinear functions
$C_{ij}$ are given by:
\begin{eqnarray}
\label{eq:collinear_factor_ij}
C_{ij}&=&-\left(\frac{\alpha_sN}{2\pi}\right)
\frac{1}{\Gamma(1-\epsilon)}\left(\frac{4\pi\mu^2}{s_{min}}\right)^\epsilon
\left\{\left[\ln\left(\frac{s_{ij}}{s_{min}}\right)-\frac{3}{4}\right]
\frac{1}{\epsilon}+\frac{1}{2}\ln^2\left(\frac{s_{ij}}{s_{min}}\right)
\right.\\
&&\left.+\frac{\pi^2}{6}-\frac{7}{4}+O(\epsilon)\right]\,\,\,,\nonumber
\end{eqnarray}
while when both $i,j=q,\bar q$, i.e. when both quarks are massless,
\begin{eqnarray}
\label{eq:collinear_factor_qq}
C_{q\bar q}&=&-\left(\frac{\alpha_sN}{2\pi}\right)
\frac{1}{\Gamma(1-\epsilon)}\left(\frac{4\pi\mu^2}{s_{min}}\right)^\epsilon
\left\{\left[2\,\ln\left(\frac{s}{s_{min}}\right)-\frac{3}{2}\right]
\frac{1}{\epsilon}+\ln^2\left(\frac{s}{s_{min}}\right)
\right.\\
&&\left.+\frac{\pi^2}{3}-\frac{7}{2}+O(\epsilon)\right]\,\,\,.\nonumber
\end{eqnarray}
Using these results, we can finally explicitly write the partonic
cross section for collinear gluon radiation as follows:
\begin{eqnarray}
\label{eq:a2_coll_tot}
\hat \sigma^{h\to q\bar qt\bar tg}_{coll}&=&
\left(\frac{\alpha_s}{2\pi}\right)
\frac{1}{\Gamma(1-\epsilon)}\left(\frac{4\pi\mu^2}{m_t^2}\right)^\epsilon
\int d(PS_4)\overline{\sum}|{\cal A}^{h\to q\bar qt\bar t}_{\sss LO}|^2
\left\{{X_{-1}^c \over \epsilon}
+N C^c_1+\frac{C^c_2}{N} \right\}\,\,\,,
\end{eqnarray}
where
\begin{eqnarray}
X_{-1}^c &=&N\left[\frac{3}{2}-
\ln\left(\frac{s_{q\bar t}}{s_{min}}\right)-
\ln\left(\frac{s_{\bar qt}}{s_{min}}\right)\right]\\
&&+\frac{1}{N}\left[-\frac{3}{2}+
2\ln\left(\frac{s}{s_{min}}\right)-
2\ln\left(\frac{s_{qt}s_{\bar q\bar t}}{s_{q\bar t}s_{\bar qt}}\right)
\right] \; ,\nonumber\\
C^c_1&=&-\ln\left(\frac{m_t^2}{s_{min}}\right)\left(
\ln\left(\frac{s_{q\bar t}}{s_{min}}\right)+
\ln\left(\frac{s_{\bar qt}}{s_{min}}\right)-\frac{3}{2}\right)
\nonumber\\
&&\,\,\,\,\,\,
-\frac{1}{2}\ln^2\left(\frac{s_{q\bar t}}{s_{min}}\right)
-\frac{1}{2}\ln^2\left(\frac{s_{\bar qt}}{s_{min}}\right)+
\frac{7}{2}-\frac{\pi^2}{3}\,\,\,,\nonumber\\
C^c_2&=&
\ln^2\left(\frac{s}{s_{min}}\right)-
\ln^2\left(\frac{s_{qt}}{s_{min}}\right)+
\ln^2\left(\frac{s_{q\bar t}}{s_{min}}\right)+
\ln^2\left(\frac{s_{\bar qt}}{s_{min}}\right)-
\ln^2\left(\frac{s_{\bar q\bar t}}{s_{min}}\right)
\nonumber\\
&&\,\,\,\,\,\,+\ln\left(\frac{m_t^2}{s_{min}}\right)
\left(-\frac{3}{2}+2\ln\left(\frac{s}{s_{min}}\right)-
2\ln\left(\frac{s_{qt}s_{\bar q\bar t}}{s_{q\bar t}s_{\bar qt}}\right)
\right)+\frac{\pi^2}{3}-\frac{7}{2}\,\,\,.\nonumber
\end{eqnarray}

\subsubsection{IR-singular gluon emission: complete result for 
$\hat\sigma_{ir}$}
\label{subsubsec:one_cutoff_ir}

As already described in the beginning of Sec.~\ref{subsec:one_cutoff},
the partonic cross section for the IR-singular real gluon radiation for
the process $q\bar q\to t\bar th$ using the one-cutoff PSS method is
given by
\begin{eqnarray}
\hat\sigma_{ir}&=& \hat \sigma_{ir}^c + \hat \sigma_{crossing} 
\nonumber \\
&=&\left[\hat \sigma^{h\to q\bar qt\bar tg}_{soft}+
\hat \sigma^{h\to q\bar qt\bar tg}_{coll}\right]_{crossed}
+\hat \sigma_{crossing} \; .
\end{eqnarray}
Note that crossing $\hat \sigma^{h\to q\bar qt\bar tg}_{soft}$ and
$\hat \sigma^{h\to q\bar qt\bar tg}_{coll}$ only implies the
interchange of the momenta of the quark and antiquark, since particle
and antiparticle interchange under crossing.  In the case of soft
gluon emission this can be easily verified by comparing
Eq.~(\ref{eq:m2_soft_lim}) with Eq.~(\ref{eq:a2_soft_fab}), after
flipping helicities and momenta of the crossed particles.  For
collinear gluon emission, the crossing is complicated by the
difference between initial and final state collinear radiation. Using
$\hat \sigma^{h\to q\bar qt\bar tg}_{soft,coll}$ in
Eqs.~(\ref{eq:a2_soft_tot}) and (\ref{eq:a2_coll_tot}), $\hat
\sigma_{ir}^c$ can be explicitly written as:
\begin{eqnarray}  
\label{eq:a2_soft_coll}
\hat\sigma_{ir}^c=
\left(\frac{\alpha_s }{2\pi}\right){\cal N}_t
\int d(PS_3)\overline{\sum}|{\cal A}_{\sss LO}|^2
\left\{
{X_{-2}^{ir}\over \epsilon^2}+{X_{-1}^{ir}\over \epsilon}
+C_1^{ir} N +C_2^{ir} {1\over N}\right\}\,\,\,,
\end{eqnarray}
where
\begin{eqnarray}
\label{eq:sigma_soft_coll_coeff}
X_{-2}^{ir}&=& -X_{-2}^{virt}\,\,\,, \\
X_{-1}^{ir}&=& -X_{-1}^{virt} \,\,\,,\nonumber\\
C_1^{ir}&=&\ln\biggl({m_t^2\over s_{min}}\biggr)\biggl[
        -2\ln\biggl({s_{\bar q\bar t}\over m_t^2}\biggr)
        -2\ln\biggl({s_{qt}\over m_t^2}\biggr)
        +{7\over 2} -\ln\biggl({m_t^2\over s_{min}}\biggr)\biggr]
\nonumber \\ &&
+\ln\biggl({s_{\bar q\bar t}\over m_t^2}\biggr)
+\ln\biggl({s_{qt}\over m_t^2}\biggr)
-\ln^2\biggl({s_{\bar q\bar t}\over m_t^2}\biggr)
-\ln^2\biggl({s_{qt}\over m_t^2}\biggr)
\nonumber \\ &&
+{7\over 2} -{\pi^2\over 2}
-\zeta(2)+m_t^2\biggl({1\over s_{\bar q\bar t}}+{1\over s_{qt}}\biggr)
\,\,\,,\nonumber \\
C_2^{ir}&=& \ln\biggl({m_t^2\over s_{min}}\biggr)\biggl[
2\ln\biggl({s \over m_t^2}\biggr)
+4\ln\biggl(\frac{s_{qt}s_{\bar q\bar t}}{s_{q\bar t}s_{\bar qt}}\biggr)
-{5\over 2}+\ln\biggl({m_t^2\over s_{min}}\biggr)
+{s_{t\bar t}\over (2m_t^2+s_{t\bar t}) 
\beta_{t\bar t}}\Lambda_{t\bar t}\biggr]
\nonumber \\ &&
+{1\over 2}\ln^2\biggl({s\over m_t^2}\biggr)
-2\ln^2\biggl({s_{\bar qt}\over m_t^2}\biggr)
+2\ln^2\biggl({s_{\bar q\bar t}\over m_t^2}\biggr)
+2\ln^2\biggl({s_{qt}\over m_t^2}\biggr)
-2\ln^2\biggl({s_{q\bar t}\over m_t^2}\biggr)
\nonumber \\
&&-2 \ln\left(\frac{s_{qt}s_{\bar q\bar t}}
{s_{\bar qt}s_{q\bar t}}\right)-
2 m_t^2\left(\frac{1}{s_{qt}}-\frac{1}{s_{q\bar t}}-
\frac{1}{s_{\bar qt}}+\frac{1}{s_{\bar q\bar t}}\right)\nonumber\\
&&
+{\pi^2\over 2}-{7\over 2}-{m_t^2\over\sqrt{\lambda_{t\bar t}}}
\left(J_a+J_b\right)\nonumber\; .
\end{eqnarray}  
while ${\cal N}_t$ is defined in Eq.~(\ref{eq:nsnt}), and ${\cal
A}_{LO}$ is the tree-level amplitude for $q\bar q\to t\bar t h$ in
$d\!=\!4$ dimensions.

As described in detail in Ref.~\cite{Giele:1993dj}, $\hat
\sigma_{crossing}$ is given by
\begin{equation}
\hat\sigma_{crossing}=\alpha_s\,\int_0^1 dz \, \hat \sigma^{q\bar q}_{\sss LO} 
\left(X_{q\to q}(z)+X_{\bar q\to\bar q}(z)\right) \; , 
\end{equation}
where $X_{q\to q}(z)$ ($X_{\bar q\to\bar q}(z)$) is the unrenormalized
crossing function of Ref.~\cite{Giele:1993dj}, which accounts for the
difference between collinear gluon radiation off an initial or a final
state quark (antiquark):
\begin{eqnarray}
X_{q\to q}(z) &=&-\frac{C_F}{2\pi}
\left(\frac{4\pi \mu^2}{s_{min}}\right)^{\epsilon} \frac{1}{\Gamma(1-\epsilon)}
\left(\frac{1}{\epsilon}\right)\times
\nonumber\\
&& \left\{\left[\frac{3}{2}-
\epsilon\left(\frac{\pi^2}{3}-\frac{7}{2}\right)\right]\delta(1-z)+
\left[\frac{1+z^2}{[(1-z)^{1+\epsilon}]_{+}}-
\epsilon (1-z)^{1-\epsilon}\right]\right\} \,\,\,.
\end{eqnarray}

\section{Total cross section for $p\bar p\rightarrow t\bar th$ 
and mass factorization}
\label{sec:total}
As described in Sec.~\ref{sec:framework}, the observable total cross
section at NLO is obtained by convoluting the parton cross section
with the NLO quark distribution functions ${\cal F}_q^{p,\bar
  p}(x,\mu)$, thereby absorbing the remaining initial-state
singularities of $\delta\hat\sigma_{\sss NLO}^{q\bar q}$ into the
quark distribution functions.  This can be understood as follows.
First the parton cross section is convoluted with the {\em bare} quark
distribution functions ${\cal F}_q^{p,\bar p}(x)$ and subsequently
${\cal F}_q^{p,\bar p}(x)$ is replaced by the renormalized quark
distribution functions ${\cal F}_q^{p,\bar p}(x,\mu)$ defined in some
subtraction scheme. Using the ${\overline{MS}}$ scheme, the
scale-dependent NLO quark distribution functions are given in terms of
${\cal F}_q^{p,\bar p}(x)$ and the QCD NLO parton distribution
function counterterms
\cite{Harris:2001sx,Giele:1993dj} as follows:\\

\underline{two-cutoff PSS method}
\begin{eqnarray}
\label{eq:pdf_mu2}
{\cal F}_q^{p,\bar p}(x,\mu)&=&
{\cal F}_q^{p,\bar p}(x) \left[1- 
\frac{\alpha_s}{2\pi}
\frac{\Gamma(1-\epsilon)}{\Gamma(1-2\epsilon)}
\left(4 \pi \right)^\epsilon
\left(\frac{1}{\epsilon}\right) C_F
\left(2\ln(\delta_s)+\frac{3}{2}\right)\right] \\
&+&\left[\frac{\alpha_s}{2\pi}
\frac{\Gamma(1-\epsilon)}{\Gamma(1-2\epsilon)}
\left(4\pi\right)^\epsilon\right]
\int_{x}^{1-\delta_s} \frac{dz}{z}
\left(-\frac{1}{\epsilon}\right) P_{qq}(z){\cal F}_j^{p,\bar p}
(\frac{x}{z})\,\,\,,\nonumber
\end{eqnarray}

\underline{one-cutoff PSS method}
\begin{eqnarray}
\label{eq:pdf_mu1}
{\cal F}_q^{p,\bar p}(x,\mu)&=&
{\cal F}_q^{p,\bar p}(x) \left[1- 
\frac{\alpha_s}{2\pi}
\frac{\left(4 \pi \right)^\epsilon}{\Gamma(1-\epsilon)}
\left(\frac{1}{\epsilon}\right) C_F \frac{3}{2} \right]\\
&+&\left[\frac{\alpha_s}{2\pi}
\frac{\left(4 \pi \right)^\epsilon}{\Gamma(1-\epsilon)}\right]
\int_{x}^{1} \frac{dz}{z}
\left(-\frac{1}{\epsilon}\right) C_F \frac{1+z^2}{(1-z)_{+}} 
{\cal F}_j^{p,\bar p}(\frac{x}{z})\,\,\,,\nonumber
\end{eqnarray}
where the ${\cal O}(\alpha_s)$ terms in the previous equations are
calculated from the ${\cal O}(\alpha_s)$ corrections to the
$q\rightarrow qg$ splitting, in the PSS formalism, and $P_{qq}(z)$ is
the Altarelli-Parisi splitting function of Eq.~(\ref{eq:p_qq}). Note
that, again, we choose the factorization and renormalization scales to
be equal. Therefore there is no explicit factorization scale
dependence in Eqs.~(\ref{eq:pdf_mu2}) and (\ref{eq:pdf_mu1}), and the
only $\mu$-dependence in ${\cal F}_q^{p,\bar p}(x,\mu)$ comes from
$\alpha_s(\mu)$. When using the two-cutoff method and convoluting the
parton cross section with the renormalized quark distribution function
of Eq.~(\ref{eq:pdf_mu2}), the IR singular counterterm of
Eq.~(\ref{eq:pdf_mu2}) exactly cancels the remaining IR poles of
$\hat{\sigma}_{virt}^{q \bar q}+\hat\sigma_{soft}$ and
$\hat{\sigma}_{hard/coll}$.  In case of the one-cutoff PSS method, the
IR singular counterterm of Eq.~(\ref{eq:pdf_mu1}) exactly cancels the
IR poles of $\hat \sigma_{crossing}$.  Finally, the complete ${\cal
  O}(\alpha_s^3)$ inclusive total cross section for $p\bar p \to t\bar
t h$ in the ${\overline{MS}}$ factorization scheme can be written as follows:\\
\begin{figure}[htbp]
\begin{center}
\hspace*{-1.5cm}
\epsfxsize=9cm \epsfbox{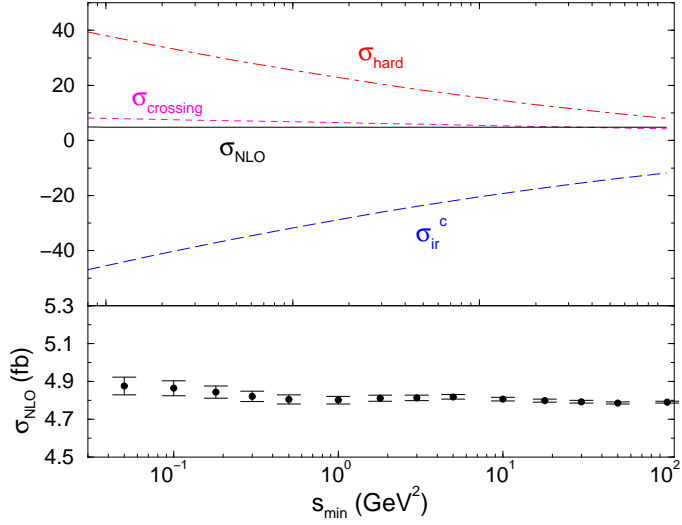}
\caption[ ]{ Dependence of $\sigma_{\sss NLO}(p {\overline p}\rightarrow
  t {\overline t} h)$ on the arbitrary cutoff of the one-cutoff PSS
  method, $s_{min}$, at $\sqrt{s_{\sss H}}\!=\!2$~TeV, for
  $M_h=120$~GeV, and $\mu=m_t$. The upper plot shows the cancellation
  of the $s_{min}$ dependence between $\sigma_{ir}^c$,
  $\sigma_{crossing}$, and $\sigma_{hard}$. The lower plot shows, on an
  enlarged scale, the dependence of $\sigma_{\sss NLO}$ on $s_{min}$,
  with the corresponding statistical errors.}
\label{fg:smin}
\end{center}
\end{figure}

\underline{two-cutoff PSS method}
\begin{eqnarray}
\label{eq:sigmatot2}
\sigma_{\sss NLO} &=&\sum_{q \bar q} \int dx_1 dx_2 {\cal F}_q^p(x_1,\mu)
{\cal F}_{\bar q}^{\bar p}(x_2,\mu) \left[
\hat{\sigma}^{q\bar q}_{\sss LO}(x_1,x_2,\mu)+
\hat{\sigma}_{virt}^{q\bar q}(x_1,x_2,\mu)+
\hat \sigma_{soft}^{\prime}(x_1,x_2,\mu)\right]\nonumber\\
&+&\frac{\alpha_s}{2\pi} C_F \sum_{q \bar q} \int dx_1 dx_2 \left\{
\int_{x_1}^{1-\delta_s}\frac{dz}{z}
\left[{\cal F}_q^p(\frac{x_1}{z},\mu) {\cal F}_{\bar q}^{\bar p}(x_2,\mu)+
{\cal F}_q^{\bar p}(x_2,\mu) {\cal F}_{\bar q}^p(\frac{x_1}{z},\mu)\right]
\right. \\
&&\times \left. \hat{\sigma}_{\sss LO}^{q\bar q}(x_1, x_2,\mu)
\left[\frac{1+z^2}{1-z}
\ln\left(\frac{s}{\mu^2}\frac{(1-z)^2}{z}\frac{\delta_c}{2}\right)
+1-z\right]+(1\leftrightarrow 2) \right\}\nonumber\\
&+& \sum_{q \bar q} \int dx_1 dx_2 {\cal F}_q^p(x_1,\mu)
{\cal F}_{\bar q}^{\bar p}(x_2,\mu) \, 
\hat{\sigma}_{hard/non-coll}(x_1,x_2,\mu) \,\,\, ,\nonumber
\end{eqnarray}
with 
\begin{equation}
\hat \sigma_{soft}^{\prime}=\hat \sigma_{soft}+
\frac{\alpha_s}{2\pi}
\frac{\Gamma(1-\epsilon)}{\Gamma(1-2\epsilon)}
\left(4 \pi \right)^\epsilon
\left(\frac{1}{\epsilon}\right) C_F
\left[4\ln(\delta_s)+3\right]\,\,\, ,
\end{equation}
\underline{one-cutoff PSS method}
\begin{eqnarray}
\label{eq:sigmatot1}
\sigma_{\sss NLO} &=&\sum_{q \bar q} \int dx_1 dx_2 {\cal F}_q^p(x_1,\mu)
{\cal F}_{\bar q}^{\bar p}(x_2,\mu) \left\{\phantom{\frac{1}{2}}
\hat{\sigma}^{q\bar q}_{\sss LO}(x_1,x_2,\mu)+
\hat{\sigma}_{virt}^{q\bar q}(x_1,x_2,\mu) \right.\\
&& \left. +\hat \sigma_{ir}^c(x_1,x_2,\mu)+
\frac{\alpha_s}{2\pi}\,2\,C_F\,\hat{\sigma}_{\sss LO}^{q\bar q}(x_1, x_2,\mu)
\left[\frac{3}{2} \ln\left(\frac{s_{min}}{\mu^2}\right) +
\frac{\pi^2}{3}-\frac{7}{2}\right]\right\}
\nonumber \\
&+&\frac{\alpha_s}{2\pi} C_F \sum_{q \bar q} \int dx_1 dx_2 \left\{
\int_{x_1}^{1}\frac{dz}{z}
\left[{\cal F}_q^p(\frac{x_1}{z},\mu) {\cal F}_{\bar q}^{\bar p}(x_2,\mu)+
{\cal F}_q^{\bar p}(x_2,\mu) {\cal F}_{\bar q}^p(\frac{x_1}{z},\mu)\right]
 \right. \nonumber \\
&&\times \left. \hat{\sigma}_{\sss LO}^{q\bar q}(x_1, x_2,\mu) 
\left[\frac{1+z^2}{(1-z)_{+}}
\ln\left(\frac{s}{\mu^2} \frac{s_{min}}{s}\right)
+1-z +(1+z^2)\left(\frac{\ln\left(1-z\right)}{1-z}\right)_{+}\right]
\right. \nonumber \\
&&\left. +
(1\leftrightarrow 2) \phantom{\frac{1}{2}}\right\}
+ \sum_{q \bar q} \int dx_1 dx_2 {\cal F}_q^p(x_1,\mu)
{\cal F}_{\bar q}^{\bar p}(x_2,\mu) \, 
\hat{\sigma}_{hard}(x_1,x_2,\mu) \,\,\, .\nonumber
\end{eqnarray}
We note that $\sigma_{\sss NLO}$ is finite, since, after mass
factorization, both soft and collinear singularities have been
canceled between $\hat{\sigma}_{virt}^{q\bar
  q}+\hat\sigma_{soft}^\prime$ and $\hat{\sigma}_{hard/coll}$ in the
two-cutoff PSS method, and between $\hat{\sigma}_{virt}^{q\bar q}$ and
$\hat\sigma_{ir}^c$ in the one-cutoff PSS method.  The last terms
respectively describe the finite real gluon emission of
Eq.~(\ref{eq:hardsplit}) and Eq.~(\ref{eq:sigma_real_one_cutoff}).
Note that the second term in Eqs.~(\ref{eq:sigmatot2}) and
(\ref{eq:sigmatot1}), which is proportional to
$\ln\left(\frac{s}{\mu^2}\right)$, corresponds exactly to the second
and third terms of Eq.~(\ref{eq:mudep_coeff}), as predicted by
renormalization group arguments.
\begin{figure}[htbp]
\begin{center}
\hspace*{-1.5cm}
\epsfxsize=9cm \epsfbox{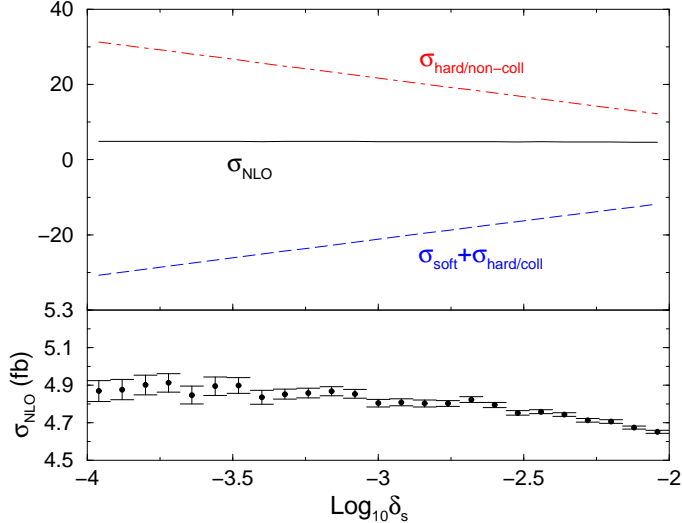}
\caption[ ]{ Dependence of $\sigma_{\sss NLO}(p {\overline p}\rightarrow
  t {\overline t} h)$ on the soft cutoff $\delta_s$ of the two-cutoff
  PSS method, at $\sqrt{s_{\sss H}}\!=\!2$~TeV, for $M_h=120$ GeV,
  $\mu=m_t$, and $\delta_c=10^{-4}$. The upper plot shows the
  cancellation of the $\delta_s$-dependence between
  $\sigma_{soft}+\sigma_{hard/coll}$ and $\sigma_{hard/non-coll}$. The
  lower plot shows, on an enlarged scale, the dependence of
  $\sigma_{NLO}$ on $\delta_s$ with the corresponding statistical
  errors.}
\label{fg:deltas}
\end{center}
\end{figure}
Before we discuss in detail the numerical results for the NLO total
cross section for $p\bar p \to t\bar th$ we first demonstrate that
$\sigma_{\sss NLO}$ does not depend on the arbitrary cutoffs of the
PSS method, i.e. on $s_{min}$ when we use the one-cutoff method, and
on the soft and hard/collinear cutoffs $\delta_s$ and $\delta_c$ when
we use the two-cutoff method.  We note that the cancellation of the
cutoff dependence at the level of the total NLO cross section is a
very delicate issue, since it involves both analytical and numerical
contributions. It is crucial to study the behavior of $\sigma_{NLO}$
in a region where the cutoff(s) are small enough to justify the
approximations used in the analytical calculation of the IR-divergent
part of $\hat\sigma_{real}^{q\bar q}$, but not so small to give origin
to numerical instabilities.  

Fig.~\ref{fg:smin} is about the one-cutoff PSS method and shows the
dependence of $\sigma_{\sss NLO}$ on $s_{min}$.  In the upper window
we illustrate the cancellation of the $s_{min}$ dependence between
$\sigma_{ir}^c$, $\sigma_{crossing}$, and $\sigma_{hard}$, while in
the lower window we show, on a larger scale, the behavior of
${\sigma}_{\sss NLO}$ including the statistical errors from the
Monte-Carlo integration. We note that $\sigma_{\sss NLO}$ also
includes the Born cross section and the virtual contribution to the
NLO cross section, which are both $s_{min}$-independent, and are
therefore not shown explicitly in the upper part of
Fig.~\ref{fg:smin}.  Clearly a plateau is reached in the region $0.1
\, \mathrm{GeV}^2 <s_{min}<100 \, \mathrm{GeV}^2$.
\begin{figure}[htbp]
\begin{center}
\hspace*{-1.5cm}
\epsfxsize=9cm \epsfbox{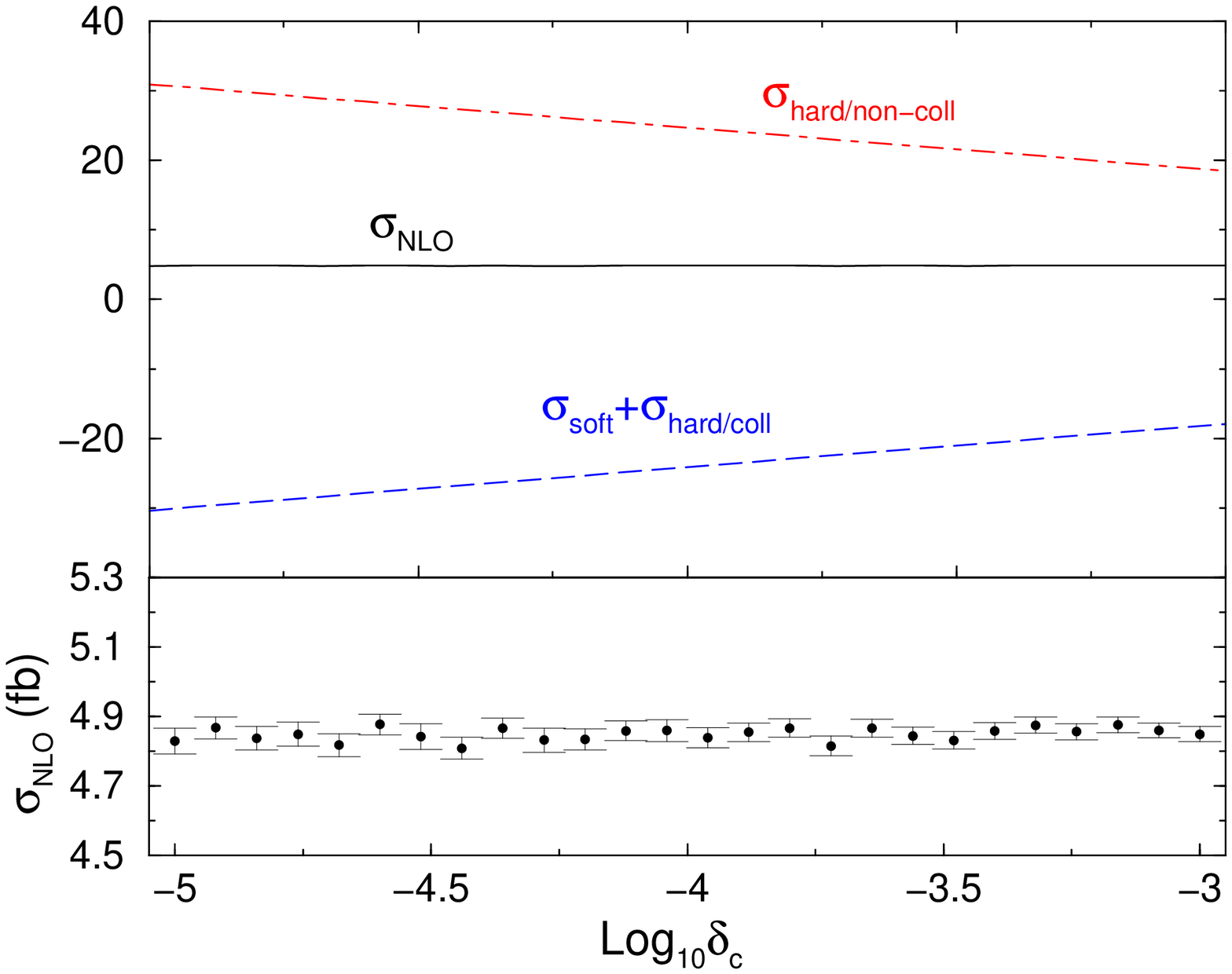}
\caption[ ]{ Dependence of $\sigma_{\sss NLO}(p {\overline p}\rightarrow
  t {\overline t} h)$ on the collinear cutoff $\delta_c$ of the
  two-cutoff PSS method, at $\sqrt{s_{\sss H}}\!=\!2$~TeV, for
  $M_h=120$ GeV, $\mu=m_t$, and $\delta_s=5\times10^{-4}$. The upper
  plot shows the cancellation of the $\delta_c$-dependence between
  $\sigma_{soft}+\sigma_{hard/coll}$ and $\sigma_{hard/non-coll}$. The
  lower plot shows, on an enlarged scale, the dependence of
  $\sigma_{NLO}$ on $\delta_c$ with the corresponding statistical
  errors.}
\label{fg:deltac}
\end{center}
\end{figure}

Figs.~\ref{fg:deltas} and \ref{fg:deltac} are about the two-cutoff PSS
method. In Fig.~\ref{fg:deltas}, we show the dependence of
$\sigma_{\sss NLO}$ on the soft cutoff, $\delta_s$, for a fixed value
of the hard/collinear cutoff, $\delta_c\!=\!10^{-4}$.  In
Fig.~\ref{fg:deltac}, we show the dependence of $\sigma_{\sss NLO}$ on
the hard/collinear cutoff, $\delta_c$, for a fixed value of the soft
cutoff, $\delta_s\!=\!5\times 10^{-4}$.  In the upper window of
Fig.~\ref{fg:deltas}(\ref{fg:deltac}) we illustrate the cancellation
of the $\delta_s$($\delta_c$) dependence between
${\sigma}_{soft}+{\sigma}_{hard/coll}$ and ${\sigma}_{hard/non-coll}$,
while in the lower window we show, on a larger scale, ${\sigma}_{\sss
  NLO}$ with the statistical errors from the Monte-Carlo integration.
As before, $\sigma_{\sss NLO}$ also includes the contribution from the
Born and the virtual cross sections, which are both cutoff-independent
and are not shown explicitly in the upper parts of
Figs.~\ref{fg:deltas},\ref{fg:deltac}.  For $\delta_s$ in the range
$10^{-4}-2.5\times10^{-3}$ and $\delta_c$ in the range
$10^{-5}-10^{-3}$, a clear plateau is reached and the NLO total cross
section is independent of the technical cutoffs of the two-cutoff PSS
method.  All the results presented in the following are obtained using
the two-cutoff PSS method with $\delta_s$ and $\delta_c$ in the range
$10^{-4}\!-\!10^{-3}$. We have confirmed them using the one-cutoff PSS
method with $1\!\le\!s_{min}\!\le\!10$.

\section{Numerical results}
\label{sec:results}

In the following we discuss in detail our results for the
NLO inclusive total cross section for 
$p {\overline p} \rightarrow t {\overline t} h$, 
$\sigma_{\sss NLO}(p {\overline p}\rightarrow t {\overline t} h)$,
as introduced in Sect.~\ref{sec:framework} and explicitly 
given by Eqs.~(\ref{eq:sigmatot2}) and (\ref{eq:sigmatot1}). 
Our numerical results are found using CTEQ4M parton distribution
functions ~\cite{Lai:1997mg} and the 2-loop evolution of
$\alpha_s(\mu)$ for the calculation of the NLO cross section, and
CTEQ4L parton distribution functions and the 1-loop evolution of
$\alpha_s(\mu)$ for the calculation of the lowest order cross section,
unless stated otherwise.  The top-quark mass is taken to be
$m_t\!=\!174$~GeV and $\alpha_s^{\sss NLO}(M_Z)\!=\!0.116$.
\begin{figure}[hbtp]
\begin{center}
\hspace*{-1.5cm}
\epsfxsize=9cm \epsfbox{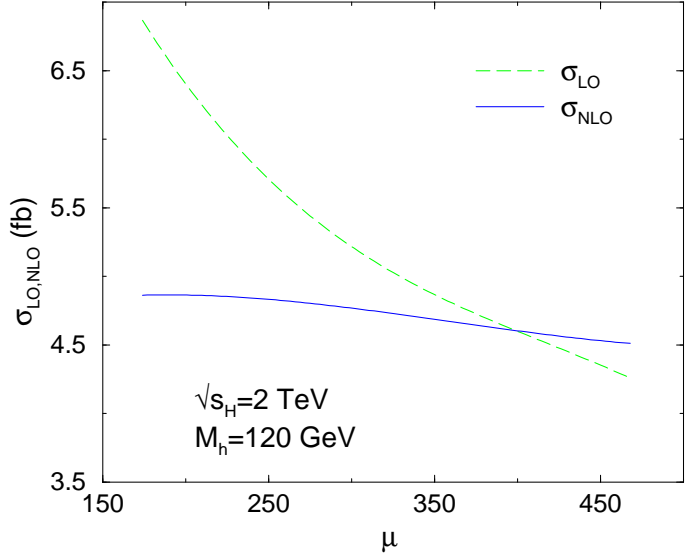}
\caption[ ]{ Dependence of $\sigma_{\sss LO,NLO}(p {\overline p}\rightarrow
  t {\overline t} h)$ on the renormalization/factorization scale $\mu$, at
  $\sqrt{s_{\sss H}}\!=\!2$~TeV, for $M_h\!=\!120$ GeV.
  }
\label{fg:mudep}
\end{center}
\end{figure}
 
First of all, in Fig.~\ref{fg:mudep} we show how at NLO the dependence
on the arbitrary renormalization/factorization scale $\mu$ is
significantly reduced. We use $M_h\!=\!120$~GeV for illustration
purposes. We note that only for scales $\mu$ of the order of $2
m_t+M_h$ or bigger is the NLO result greater than the lowest order
result at $\sqrt{s_{\sss H}}=2$~TeV.
\begin{figure}[htbp]
\begin{center}
\hspace*{-1.5cm}
\epsfxsize=9cm \epsfbox{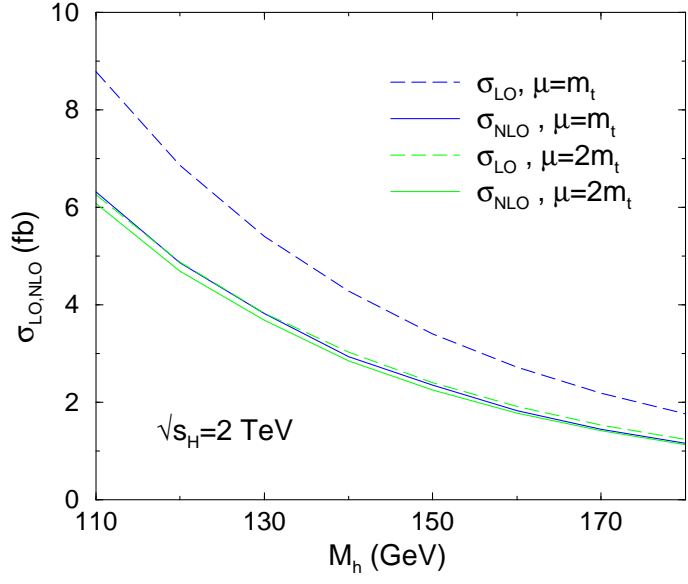}
\caption[ ]{$\sigma_{\sss NLO}$ and $\sigma_{\sss LO}$  for
  $p {\overline p} \rightarrow t {\overline t} h$ as functions of
  $M_h$, at $\sqrt{s_{\sss H}}\!=\!2$~TeV, for
  $\mu\!=m_t$ and $\mu\!=\!2m_t$.}
\label{fg:signlo}
\end{center}
\end{figure}
\begin{figure}[htbp]
\begin{center}
\hspace*{-1.5cm}
\epsfxsize=9cm \epsfbox{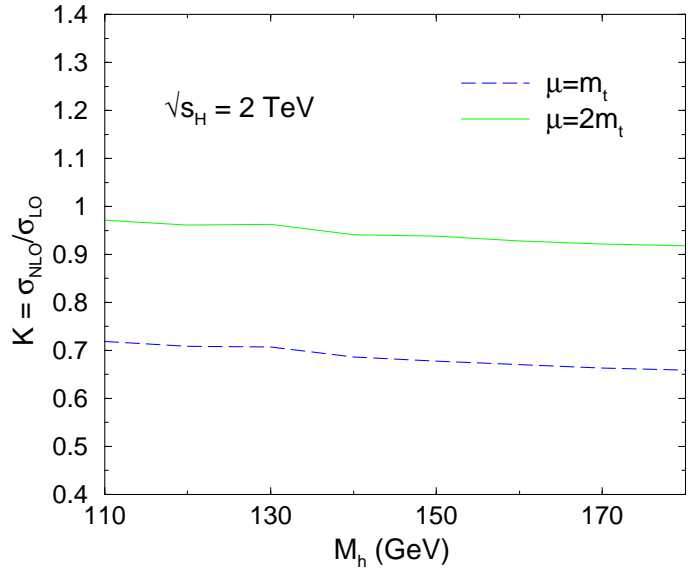}
\caption[ ]{K-factor for $p
  {\overline p}\rightarrow t {\overline t} h$ as a function of $M_h$,
  at $\sqrt{s_{\sss H}}\!=\!2$~TeV, for $\mu\!=m_t$ and
  $\mu\!=\!2m_t$ .}
\label{fg:sigk}
\end{center}
\end{figure}

Fig.~\ref{fg:signlo} shows both the LO and the NLO total cross section
for $p\bar p\rightarrow t\bar th$ as a function of $M_h$, at
$\sqrt{s_{\sss H}}\!=\!2$~TeV, for two values of the
renormalization/factorization scale, $\mu\!=\!m_t$ and $\mu\!=\!2
m_t$.  Over the entire range of $M_h$ accessible at the Tevatron, the
NLO corrections decrease the rate for renormalization/factorization
scales $\mu < 2 m_t+M_h$.  The reduction is much less dramatic at
$\mu\!=\!2m_t$ than at $\mu\!=\!m_t$, as can be seen from both
Fig.~\ref{fg:mudep} and Fig.~\ref{fg:signlo}. An illustrative sample
of results is also given in Table~1. The error we quote on our values
is the statistical error of the numerical integration involved in
evaluating the total cross section.  We estimate the remaining
theoretical uncertainty on the NLO results to be of the order of
$12\%$. This is mainly due to: the left over $\mu$-dependence (about
$8\%$), the dependence on the PDFs (about $6\%$), and the error on
$m_t$ (about $7\%$) which particularly plays a role in the Yukawa
coupling.
\begin{table}
\begin{center}
\begin{tabular}{|c|c|c|c|c|}
\hline\hline
$M_h$ (GeV) & $\mu$ & $\sigma_{\sss LO}$ (fb) &$\bar\sigma_{\sss LO}$ (fb)
& $\sigma_{\sss NLO}$ (fb) \\
\hline\hline
    & $m_t$       & 6.8662 $\pm$ 0.0013 & 5.2843 $\pm$ 0.0008 
                  & 4.863 $\pm$  0.029 \\
120 & $m_t+M_h/2$ & 5.9085 $\pm$ 0.0011 & 4.5846 $\pm$ 0.0007
                  & 4.847 $\pm$  0.024 \\
    & $2m_t$      & 4.8789 $\pm$ 0.0009 & 3.8252 $\pm$ 0.0006
                  & 4.691 $\pm$  0.020 \\
    & $2m_t+M_h$  & 4.2548 $\pm$ 0.0008 & 3.3600 $\pm$ 0.0005
                  & 4.511  $\pm$ 0.017 \\\hline
    & $m_t$       & 3.4040 $\pm$ 0.0006 & 2.5811 $\pm$ 0.0005
                  & 2.355 $\pm$ 0.013  \\
150 & $m_t+M_h/2$ & 2.8289 $\pm$ 0.0005 & 2.1668 $\pm$ 0.0004
                  & 2.315 $\pm$ 0.011 \\
    & $2m_t$      & 2.4007 $\pm$ 0.0004 & 1.8553 $\pm$ 0.0004
                  & 2.253 $\pm$ 0.010 \\
    & $2m_t+M_h$  & 2.0282 $\pm$ 0.0004 & 1.5813 $\pm$ 0.0003
                  & 2.147 $\pm$ 0.008 \\\hline
    & $m_t$       & 1.7605 $\pm$ 0.0003 & 1.3153 $\pm$ 0.0002
                  & 1.160  $\pm$ 0.007 \\
180 & $m_t+M_h/2$ & 1.4142 $\pm$ 0.0003 & 1.0693 $\pm$ 0.0002
                  & 1.158 $\pm$ 0.005 \\
    & $2m_t$      & 1.2326 $\pm$ 0.0002 & 0.9390 $\pm$ 0.0001
                  &1.132 $\pm$ 0.004 \\
    & $2m_t+M_h$  & 1.0096 $\pm$ 0.0002 & 0.7773 $\pm$ 0.0001
                  & 1.069 $\pm$ 0.004 \\
 \hline\hline
\end{tabular}
\caption{Values of both $\sigma_{\sss LO}$ (calculated with LO
  $\alpha_s(\mu)$ and LO PDFs), $\bar\sigma_{\sss LO}$ (calculated with
  NLO $\alpha_s(\mu)$ and NLO PDFs), and $\sigma_{\sss NLO}$ for
  different values of $M_h$ and for different
  renormalization/factorization scales $\mu$.}
\end{center}
\end{table}

The corresponding K-factor, i.e. the ratio of the NLO cross section to
the LO one,
\begin{equation}
K=\frac{\sigma_{\sss NLO}}{\sigma_{\sss LO}} \; ,
\end{equation}
is shown in Fig.~\ref{fg:sigk}. For scales $\mu$ between $\mu\!=\!m_t$
and $\mu\!=\!2m_t$, the K-factor varies roughly between $K\!=\!0.70$
and $K\!=\!0.95$, when $M_h$ varies in the range between $100$ and
$200$~GeV.  For scales of the order of $\mu\!=\!2m_t+M_h$ the K-factor
is of order one and becomes larger than one for higher scales.  Given
the strong scale dependence of the LO cross section, the K-factor also
shows a significant $\mu$-dependence and therefore is an equally
unreliable prediction.  Moreover it is important to remember that the
K-factor depends on how the LO cross section is calculated.  We choose
to calculate the LO cross section using both LO $\alpha_s(\mu)$ and LO
PDFs, denoted by $\sigma_{\sss LO}$ in Table~1.  An equally valid
approach could be to evaluate the LO cross section using NLO
$\alpha_s(\mu)$ and NLO PDFs, denoted by ${\bar\sigma}_{\sss LO}$ in
Table~1, in which case the K-factor would just represent the impact of
the ${\cal O}(\alpha_s)$ corrections that do not originate from the
running of $\alpha_s(\mu)$ and the PDFs. Since $\sigma_{\sss
  LO}\!>\!\bar\sigma_{\sss LO}$, the K-factor obtained using
$\sigma_{\sss LO}$ is smaller than the one obtained using
$\bar\sigma_{\sss LO}$, and it is important to match the right
K-factor to the right $\sigma_{\sss LO}$ or $\bar\sigma_{\sss LO}$.
Therefore we would like to stress once more that we only discuss the
K-factor as a qualitative indication of the impact of ${\cal
  O}(\alpha_s)$ QCD corrections, for different processes or when using
different approaches. The physical meaningful quantity is the NLO
cross section, not the K-factor.

For comparison, we have estimated the K-factor also in the EHA
\cite{Dawson:1998im}, and we obtain $K\!\simeq\!0.6-0.7$, for Higgs
boson masses up to 150 GeV and renormalization/factorization scales 
in the range between
$\mu\!=\!m_t$ and $\mu\!=\!2 m_t+M_h$. As anticipated, we do not
expect the EHA to give a quantitatively good approximation of the full
$p\bar p\rightarrow t\bar t h$ calculation at ${\cal O}(\alpha_s)$,
since at $\sqrt{s_{\sss H}}\!=\!2$ TeV and for a SM Higgs boson above the
experimental bound, we cannot work in the limit
$M_h,m_t/\sqrt{s}\!\ll\!1$ or $M_h/m_t\!\ll\!1$. Still the EHA gives a
remarkably good qualitative indication of the fact that the first
order QCD corrections may lower the LO total cross section.

It is interesting to compare our NLO result for $p\bar p\rightarrow t
{\overline t} h$ with the NLO result for $p {\overline p} \rightarrow
t {\overline t}$.  Since the Higgs boson is colorless, one would
naively expect the QCD corrections to both processes to be of roughly
the same size.  Defining the NLO cross section using the NLO evolution
of $\alpha_s(\mu)$ and the NLO CTEQ4M PDFs, and the LO cross section
using the LO evolution of $\alpha_s(\mu)$ and the LO CTEQ4L PDFs, the
K-factor for $t {\overline t}$ production at $\sqrt{s_{\sss
    H}}\!=\!2$~TeV, with $\mu\!=\!m_t$ and $m_t\!=\!174$~GeV, is:
\begin{eqnarray}
\label{eq:kfac_ttbar}
K(p {\overline p}\rightarrow t {\overline t})\mid_{q \overline q} 
&=&0.98 \,\,\,,\nonumber \\ 
K(p {\overline p}\rightarrow t {\overline t})\mid_{tot} 
&=&1.05 \; ,
\end{eqnarray}
where the $q {\overline q}$ label indicates that only the $q
{\overline q}$ initial state is included. The size of the QCD
corrections to $p {\overline p} \rightarrow t {\overline t}$ is thus
similar in magnitude to the result obtained in Fig.~\ref{fg:sigk},
taking into account that $p\bar p\rightarrow t\bar th$ is completely
dominated by the $q\bar q$ channel. Of course, we do not
expect a better agreement, since in $p\bar p\rightarrow t\bar th$ an
additional heavy particle is produced, and new contributions to the
virtual corrections arise. Moreover, taking the EHA as an indication,
one could naively expect that the radiation of a Higgs boson
introduces an additional negative contribution.  We also observe that,
if we now use as LO cross section the one obtained using NLO
$\alpha_s(\mu)$ and NLO CTEQ4M PDFs, the two K-factors in
Eq.~(\ref{eq:kfac_ttbar}) increase, according to the comment we made
above, and become:
\begin{eqnarray}
K(p {\overline p}\rightarrow t {\overline t})\mid_{q \overline q} 
&=&1.18 \,\,\,,\nonumber \\ 
K(p {\overline p}\rightarrow t {\overline t})\mid_{tot} 
&=&1.24 \;,
\end{eqnarray}
in agreement with the literature \cite{Bonciani:1998vc}\footnote{We
  have compared our results with Fig.~9 of
  Ref.~\cite{Bonciani:1998vc}, and we see very good agreement with the
  LO and the NLO curves, using $m_t\!=\!175$ GeV and $\sqrt{s_{\sss
      H}}\!=\!1.8$~TeV.}.  Moreover, since the NLO cross section for
$p\bar p\rightarrow t\bar t$ is further increased by the resummation
of the leading and next-to-leading logarithms arising from the
threshold region dynamics, the total K-factor for $p\bar p\rightarrow
t\bar t$ can be as high as 1.33 for $\mu\!=\!m_t$. With this respect,
we also note that, contrary to $p\bar p\rightarrow t\bar t$, in the
threshold region for $p\bar p\rightarrow t\bar th$ there are large
negative contributions, mainly from soft gluon radiation, which are
largely compensated by large positive contributions from hard gluon
radiation at larger $\sqrt{s}$.  In the threshold region the Coulomb
term, coming from the exchange of virtual gluons between the $t/\bar
t$ external legs, is important and contributes to decrease the NLO
cross section, although it is moderated by the behavior of the
three-body phase space.  In the strict threshold limit, the Coulomb
contribution to $p\bar p\to t\bar th$ goes to zero, while for $t \bar
t$ production it is constant and dominates the NLO cross section.

\section{Conclusion}
\label{sec:conclusion}
The NLO inclusive total cross section for the Standard Model process
$p {\overline p}\rightarrow t {\overline t} h$ at
$\sqrt{s}_H\!=\!2$~TeV shows a significantly reduced scale dependence
as compared to the Born result and leads to increased confidence in
predictions based on these results.  The NLO QCD corrections slightly
decrease or increase the Born level cross section depending on the
renormalization/factorization scales used. The NLO inclusive total
cross section for Higgs boson masses in the range accessible at the
Tevatron, $120\!<\!M_h\!<\!180$ GeV, is of the order of $1-5$ fb.
 
The contributions to the NLO cross section resulting from real gluon
emission have been calculated in two variations of the phase space
slicing method, involving one or two arbitrary numerical cutoff
parameters, respectively.  This is the first application of the 
one-cutoff phase space slicing approach, (``$s_{min}$''), to a cross
section involving more than one massive particle in the final state.
The correspondence between the two phase space slicing approaches is
made explicit.  The virtual contributions to the NLO cross section
require the calculation of both box and pentagon diagrams involving
several massive particles and explicit results for the integrals have
been presented in the appendices. These techniques can now be applied
to other similar processes.

\section*{Acknowledgments}
We are particularly thankful to Z.~Bern and F.~Paige for valuable
discussions and encouragement. We would like to thank W.~Giele,
S.~Keller, and W.~Kilgore for very useful suggestions and insights.
We are grateful to the authors of Ref.~\cite{Beenakker:2001rj} for
detailed comparisons of results prior to publication.  The work of
L.R. (S.D.)  is supported in part by the U.S. Department of Energy
under grant DE-FG02-97ER41022 (DE-AC02-76CH00016). The work of D.W.
is supported by the U.S. Department of Energy under grant
DE-FG02-91ER40685.

\appendix
\section{Pentagon scalar integrals}
\label{sec:scalar_pentagon}

In this appendix we review the details of the calculation of the 
pentagon scalar integrals that appear in the calculation of diagrams $P_1$
and $P_2$ illustrated in Fig.~\ref{fig:pentagons}. Using the momentum
flow and the notation shown in Fig.~\ref{fig:pentagons}, the pentagon 
scalar integral originating from diagram $P_1$ ($E0_{p1}$) can be 
written as:
\begin{equation}
\label{eq:p1_int}
E0_{p1}=\mu^{4-d}
\int\frac{d^dk}{(2\pi)^d}\frac{1}{N_1 N_2 N_3 N_4 N_5} \,\,\,,
\end{equation}
where
\begin{eqnarray}
\label{eq:p1_denominators} 
N_1&=&k^2 \,\,\,,\\ 
N_2&=& (k+q_1)^2\,\,\,,\nonumber \\ 
N_3&=& (k+q_1+q_2)^2\,\,\,,\nonumber \\
N_4&=& (k+q_1+q_2-p_t^\prime)^2-m_t^2\,\,\,,\nonumber\\
N_5&=& (k+q_1+q_2-p_t^\prime-p_h)^2-m_t^2\,\,\,.\nonumber
\end{eqnarray}
We note that we included a factor $\mu^{4-d}$ in the definition of the
$d$-dimensional scalar integrals in order to have them in the most
convenient form for the calculation of the virtual amplitude squared.
The pentagon scalar integral originating from diagram $P_2$,
$E0_{p2}$, can be obtained from Eqs.(\ref{eq:p1_int}) and
(\ref{eq:p1_denominators}) by exchanging $q_1\leftrightarrow q_2$.
Therefore in the following we limit our discussion to $E0_{p1}$, the
generalization to $E0_{p2}$ being straightforward.

We calculate these integrals following the method introduced by the
authors of Ref.~\cite{Bern:1993em}. To make contact with their
notation, we denote by $k_i$ the external momenta
(such that $k_i^2\!=\!m_i^2$), by $M_i$ the internal masses, by $p_i$ the
sum of the first $i$ external momenta, $p_i^\mu\!=\!\sum_{j=1}^i k_j^\mu$,
 by $p_{ij}$ the difference $p_{ij}^\mu\!=\!p_{j-1}^\mu-p_{i-1}^\mu=
k_i^\mu+k_{i+1}^\mu+\cdots+k_{j-1}^\mu$ (for $i<j$), and finally by 
${\bar s}_{ij}$ the invariant masses ${\bar s}_{ij}\!=\!(k_i+k_j)^2$.

\begin{figure}[bt]
\centering
\epsfxsize=2.in
\leavevmode\epsffile{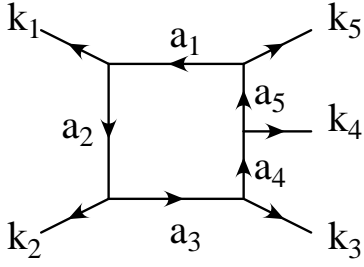}
\caption[]{Topology of the pentagon scalar integral}
\label{fig:pentagon_red}
\end{figure}

The topology of the generic pentagon scalar integral is illustrated in
Fig.~\ref{fig:pentagon_red}, which can be specified to our case by
identifying:
\begin{eqnarray}
k_1 &\longrightarrow& -q_1\,\,\,\,(\mbox{incoming}\,\, q) \nonumber\\
k_2 &\longrightarrow& -q_2\,\,\,\,(\mbox{incoming}\,\, \bar q)\nonumber\\
k_3 &\longrightarrow& p_t^\prime\,\,\,\,(\mbox{outgoing}\,\, \bar t)\\
k_4 &\longrightarrow& p_h\,\,\,\,(\mbox{outgoing}\,\, h)\nonumber\\
k_5 &\longrightarrow& p_t\,\,\,\,(\mbox{outgoing}\,\, t)\nonumber
\; .
\end{eqnarray}

Using the standard Feynman parameterization technique, the pentagon
integral in Eq.(\ref{eq:p1_int}) can be written as:
\begin{equation}
E0_{p1}=-{i\over 16\pi^2}(4\pi\mu^2)^\epsilon\Gamma(3+\epsilon)
\int_0^1{\Pi_{i=1}^5 da_i \delta(1-\Sigma_{i=1}^5 a_i)\over [{\cal D}_{p1}(a_i)
]^{3+\epsilon}}\,\,\,,
\label{eq:pent_fp}
\end{equation}
where the denominator ${\cal D}_{p1}(a_i)$ is:
\begin{equation}
{\cal D}_{p1}(a_i)=\sum_{i,j=1}^5 S_{ij}a_ia_j-i\eta\,\,\,,
\end{equation}
and the symmetric matrix $S_{ij}$ is given by:
\begin{equation}
S_{ij}=\frac{1}{2}(M_i^2+M_j^2-p_{ij}^2)\,\,\,.
\end{equation}
For our particular process, the matrix $S_{ij}$ has the following
explicit form:
\begin{equation}
\label{eq:s_matrix}
S=\frac{1}{2}\left(
\begin{array}{c c c c c}
0 & 0 & -{\bar s}_{12} & (m_t^2-{\bar s}_{45}) & 0\\
0 & 0 & 0 & (m_t^2-{\bar s}_{23}) & (m_t^2-{\bar s}_{15}) \\
-{\bar s}_{12} & 0 & 0 & 0 & (m_t^2-{\bar s}_{34}) \\
(m_t^2-{\bar s}_{45}) & (m_t^2-{\bar s}_{23}) & 0 & 2m_t^2 & (2 m_t^2-M_h^2) \\
0 & (m_t^2-{\bar s}_{15}) & (m_t^2-{\bar s}_{34}) & (2 m_t^2-M_h^2) & 2 m_t^2
\end{array}
\right)\,\,\,.
\end{equation}
Following Ref.~\cite{Bern:1993em}, $E0_{p1}$ can then be written as the
linear combination of five scalar box integrals $D0_{p1}^{(i)}$:
\begin{equation}
\label{eq:pentagon_sumofboxes}
E0_{p1}=-\frac{1}{2}\sum_{i=1}^5 c_i D0_{p1}^{(i)}\,\,\,,
\end{equation}
where each $D0_{p1}^{(i)}$ scalar box integral can be obtained from
the scalar pentagon integral $E0_{p1}$ of Eq.~(\ref{eq:pent_fp}) in
the limit where one of the Feynman parameters $a_i$ of the internal
propagators goes to zero (i.e. $D0_{p1}^{(i)}$ is obtained when
$a_i\rightarrow 0$). The five box scalar integrals we need are given
in Secs.~\ref{subsec:D0_1_p1}-\ref{subsec:D0_5_p1}.  The coefficients
$c_i$ in Eq.~(\ref{eq:pentagon_sumofboxes}) are given by:
\begin{equation}
\label{eq:pentagon_coefficients}
c_i=\sum_{j=1}^{5}S_{ij}^{-1} \,\,\,.
\end{equation}
Using Eq.~(\ref{eq:s_matrix}) we can easily obtain them in terms of
$m_t$, $M_h$, and the kinematic invariants ${\bar s}_{ij}$.

The final result for the pentagon scalar integral $E0_{p1}$ can be
written as:
\begin{equation}
E0_{p1}=\frac{i}{16\pi^2}{\cal N}_t\biggl[ {X_{-2}\over \epsilon^2}
+{X_{-1}\over \epsilon}+X_0\biggr]\,\,\,,
\end{equation}
where ${\cal N}_t$ is given in Eq.~(\ref{eq:nsnt}), while $X_{-2}$,
$X_{-1}$ and $X_0$ are obtained using
Eqs.~(\ref{eq:pentagon_sumofboxes})-(\ref{eq:s_matrix}), and the
results in Secs.~\ref{subsec:D0_1_p1}-\ref{subsec:D0_5_p1}. The
expression for $X_0$ is too lengthy to be given explicitly in this
appendix, while $X_{-2}$ and $X_{-1}$ have the following compact form:
\begin{eqnarray}
X_{-2} &=& \frac{1}{2\sigma}\left(-\frac{1}{\omega_1\tau_1}-
\frac{1}{\omega_2\tau_2}+\frac{2}{\tau_1\tau_2}\right)\,\,\,,\\
X_{-1} &=& \frac{1}{\sigma\tau_1\tau_2}\left(-\Lambda_\sigma+
\Lambda_{\omega_1}+\Lambda_{\omega_2}-\Lambda_{\tau_1}-\Lambda_{\tau_2}\right)
+\frac{1}{\sigma\tau_2\omega_2}\left(\Lambda_{\tau_2}-\Lambda_{\tau_1}+
\Lambda_{\omega_2}\right)+\nonumber\\
&+&\frac{1}{\sigma\tau_1\omega_1}\left(\Lambda_{\tau_1}-\Lambda_{\tau_2}+
\Lambda_{\omega_1}\right)\,\,\,,\nonumber
\end{eqnarray} 
where we have defined:
\begin{eqnarray}
\label{eq:den_symbols}
\sigma&=&(q_1+q_2)^2=s\,\,\,,\\
\tau_1 &=& m_t^2-(q_1-p_t)^2=2\,q_1\cdot p_t =s_{qt}\,\,\,,\nonumber\\
\tau_2 &=& m_t^2-(q_2-p_t^\prime)^2=2\,q_2\cdot p_t^\prime=
s_{\bar q\bar t}\,\,\,, \nonumber\\
\omega_1 &=& (p_t+p_h)^2-m_t^2\,\,\,, \nonumber\\
\omega_2 &=& (p_t^\prime+p_h)^2-m_t^2\,\,\,, \nonumber
\end{eqnarray}
and
\begin{eqnarray}
\Lambda_\sigma&=&\ln\left(\frac{\sigma}{m_t^2}\right)\,\,\,\,,\,\,\,\,
\Lambda_{\tau_1}=\ln\left(\frac{\tau_1}{m_t^2}\right)\,\,\,\,,\,\,\,\,
\Lambda_{\tau_2}=\ln\left(\frac{\tau_2}{m_t^2}\right)\,\,\,\,,\\
\Lambda_{\omega_1}&=&\ln\left(\frac{\omega_1}{m_t^2}\right)\,\,\,\,,\,\,\,\,
\Lambda_{\omega_2}=\ln\left(\frac{\omega_2}{m_t^2}\right)\,\,\,\,.\nonumber
\end{eqnarray}
We discuss in the following the box scalar integrals $D0_{p1}^{(i)}$,
which are used in Eq.~(\ref{eq:pentagon_sumofboxes}) to calculate
$E0_{p1}$.  The analogous box scalar integrals for $E0_{p2}$ can be
obtained from the $D0_{p1}^{(i)}$ by exchanging $q_1\leftrightarrow
q_2$ in their analytic expression.

\subsection{Box scalar integral \boldmath $D0_{p1}^{(1)}$\unboldmath}
\label{subsec:D0_1_p1}
$D0_{p1}^{(1)}$ is obtained from the pentagon in the limit
$a_1\rightarrow 0$ and corresponds to the following integral:
\begin{eqnarray}
\label{eq:D0_1_p1_integral}
D0_{p1}^{(1)}&=&\mu^{4-d}
\int\frac{d^d k}{(2\pi)^d}\frac{1}{N_2N_3N_4N_5}\\
&=&\mu^{4-d}\int\frac{d^d k}{(2\pi)^d}
\frac{1}{k^2 (k+q_2)^2[(k+q_2-p_t^\prime)^2-m_t^2][(k+q_2-p_t^\prime-p_h)^2
-m_t^2]}\,\,\,,\nonumber
\end{eqnarray}
after the momentum shift $k\rightarrow k-q_1$ has been applied to the
denominators listed in Eq.(\ref{eq:p1_denominators}). The part of 
$D0_{p1}^{(1)}$ which contributes to the virtual amplitude squared
is given by:
\begin{equation} 
\label{eq:D0_1_p1_result}
D0_{p1}^{(1)}=\frac{i}{16\pi^2}{\cal N}_t\left(-\frac{1}
{\omega_2\tau_2}\right)
\left(\frac{X_{-2}}{\epsilon^2}+\frac{X_{-1}}{\epsilon}+X_0\right)\,\,\,,
\end{equation}
where ${\cal N}_t$ is given in Eq.~(\ref{eq:nsnt}), while the
coefficients $X_{-2}$, $X_{-1}$, and $X_{0}$ are given by:
\begin{eqnarray}
\label{eq:D0_1_p1_coefficients}
X_{-2}&=&\frac{1}{2}\,\,\,, \\
X_{-1}&=& \ln\left(\frac{\tau_1 m_t^2}{\omega_2\tau_2}\right)\,\,\,,
\nonumber \\
X_0&=& Re\left\{-\frac{5}{6}\pi^2+
 \ln^2\left(\frac{\omega_2}{m_t^2}\right)+
 \ln^2\left(\frac{\tau_2}{m_t^2}\right)-
 \ln^2\left(\frac{\tau_1}{m_t^2}\right)\right.\\
 &+&2\,\ln\left(\frac{\omega_2+\tau_1}{\tau_2}\right)
    \ln\left(\frac{\tau_1}{\omega_2}\right)+
 2\,\ln\left(\frac{\tau_2-\tau_1}{\omega_2}\right)
    \ln\left(\frac{\tau_1}{\tau_2}\right)\nonumber\\
 &-&2\left.\,\mbox{Li}_2\left(\frac{\tau_2-\tau_1-\omega_2}{\tau_2}\right)-
 2\,\mbox{Li}_2\left(\frac{\omega_2+\tau_1-\tau_2}{\omega_2}\right)+
 2\,\mbox{Li}_2\left(\frac{\tau_1(\omega_2+\tau_1-\tau_2)}{\omega_2\tau_2}
\right)-
 {\cal I}_0\right\}\,\,\,,\nonumber
\end{eqnarray}
where
\begin{eqnarray}
{\cal I}_0&=& 
  \ln\left(\frac{\tau_2}{\tau_1}\right)\ln\left(\frac{M_h^2}{m_t^2}\right)+
\left\{-\mbox{Li}_2\left(\frac{1}{\lambda_+}\right)+
\ln\left(\frac{\tau_2}{\tau_1}\right)
\ln\left(\frac{-\tau_1-\lambda_+(\tau_2-\tau_1)}{\tau_2-\tau_1}\right)\right.
\nonumber\\
&-&\left.
\mbox{Li}_2\left(\frac{\tau_2}{\lambda_+(\tau_2-\tau_1)+\tau_1}\right)+
\mbox{Li}_2\left(\frac{\tau_1}{\lambda_+(\tau_2-\tau_1)+\tau_1}\right)+
(\lambda_+\leftrightarrow\lambda_-)\right\}\,\,\,,
\end{eqnarray}
and
\begin{equation}
\label{eq:D0_1_p1_lambdapm}
\lambda_\pm=\frac{1}{2}\left(1\pm \sqrt{1-\frac{4m_t^2}{M_h^2}}\right)
\,\,\,.
\end{equation}

\subsection{Box scalar integral \boldmath $D0_{p1}^{(2)}$\unboldmath}
\label{subsec:D0_2_p1}
$D0_{p1}^{(2)}$ is obtained from the pentagon in the limit $a_2\rightarrow
0$ and corresponds to the following integral:
\begin{eqnarray}
\label{eq:D0_2_p1_integral}
D0_{p1}^{(2)}&=&\mu^{4-d}
\int\frac{d^d k}{(2\pi)^d}\frac{1}{N_1N_3N_4N_5}\\
&=&\mu^{4-d}\int\frac{d^d k}{(2\pi)^d}\frac{1}
{k^2 (k+q)^2[(k+q-p_t^\prime)^2-m_t^2][(k+q-p_t^\prime-p_h)^2-m_t^2]}\,\,\,.
\nonumber
\end{eqnarray}
$D0_{p2}^{(2)}$ is equal to $D0_{p1}^{(2)}$, and they both coincide with
$D0_{b1}$ in Appendix~\ref{subsec:b1_scalar}. 

\subsection{Box scalar integral \boldmath $D0_{p1}^{(3)}$ \unboldmath}
\label{subsec:D0_3_p1}
$D0_{p1}^{(3)}$ is obtained from the pentagon in the limit
$a_3\rightarrow 0$ and corresponds to the following integral:
\begin{eqnarray}
\label{eq:D0_3_p1_integral}
D0_{p1}^{(3)}&=&\mu^{4-d}\int\frac{d^d k}{(2\pi)^d}\frac{1}{N_1N_2N_4N_5}\\
&=&\mu^{4-d}\int\frac{d^d k}{(2\pi)^d}\frac{1}
{k^2 (k+q_1)^2[(k+q_1-p_t)^2-m_t^2][(k+q_1-p_t-p_h)^2-m_t^2]}\,\,\,,
\nonumber
\end{eqnarray}
after the momentum shift $k\rightarrow-k-q_1$ has been applied to the
denominators in Eq.~(\ref{eq:p1_denominators}). We notice that this
integral can be obtained from $D0_{p1}^{(1)}$ when $q_2\rightarrow
q_1$ and $p_t^\prime\rightarrow p_t$.

\subsection{Box scalar integral \boldmath $D0_{p1}^{(4)}$\unboldmath}
\label{subsec:D0_4_p1}
$D0_{p1}^{(4)}$ is obtained from the pentagon in the limit
$a_4\rightarrow 0$ and corresponds to the following integral:
\begin{eqnarray}
\label{eq:D0_4_p1_integral}
D0_{p1}^{(4)}&=&\mu^{4-d}\int\frac{d^d k}{(2\pi)^d}\frac{1}{N_1N_2N_3N_5}\\
&=& \mu^{4-d}\int\frac{d^d k}{(2\pi)^d}\frac{1}
{k^2 (k+q_2)^2(k+q_1+q_2)^2[(k+q_1+q_2-p_t)^2-m_t^2]}\,\,\,,\nonumber
\end{eqnarray}
after the momentum shift $k\rightarrow-k-q_1-q_2$ has been applied to
the denominators in Eq.~(\ref{eq:p1_denominators}).  This integral
coincides with $D0_{b3}^{(3)}$ in Appendix~\ref{subsec:b3_scalar}.

\subsection{Box scalar integral \boldmath $D0_{p1}^{(5)}$\unboldmath}
\label{subsec:D0_5_p1}
$D0_{p1}^{(5)}$ is obtained from the pentagon in the limit
$a_5\rightarrow 0$ and corresponds to the following integral:
\begin{eqnarray}
\label{eq:D0_5_p1_integral}
D0_{p1}^{(5)}&=&\mu^{4-d}\int\frac{d^d k}{(2\pi)^d}\frac{1}{N_1N_2N_3N_4}\\
&=&\mu^{4-d}\int\frac{d^d k}{(2\pi)^d}\frac{1}
{k^2 (k+q_1)^2(k+q_1+q_2)^2[(k+q_1+q_2-p_t^\prime)^2-m_t^2]}\nonumber
\end{eqnarray}
This integral coincides with $D0_{b3}^{(1)}$ in
Appendix~\ref{subsec:b3_scalar}.
\section{Box scalar integrals}
\label{sec:scalar_boxes}
\subsection{Box 1 : box scalar integral \boldmath$D0_{b1}$\unboldmath}
\label{subsec:b1_scalar}
The scalar box integral $D0_{b1}$ can be written as:
\begin{equation} 
\label{eq:i_b1_integral}
D0_{b1}=\mu^{4-d}\int\frac{d^dk}{(2\pi)^d}\frac{1}{N_1 N_2 N_3 N_4} \,\,\,,
\end{equation} 
where 
\begin{eqnarray} 
\label{eq:i_b1_denominators}
N_1&=&k^2\,\,\,,\nonumber \\ 
N_2&=& (k+q)^2 \,\,\,,\\ 
N_3&=& (k+q-p_t^\prime)^2-m_t^2\,\,\,,\nonumber \\
N_4&=& (k+q-p_t^\prime-p_h)^2-m_t^2 \,\,\,.\nonumber
\end{eqnarray}
The analytical expression for this integral can be found in
Ref.~\cite{Denner:1991qq}. Since the integral is finite, we have
evaluated it in $d=4$ dimensions 
using the FF package \cite{vanOldenborgh:1990wn}.

\subsection{Box 2: box scalar integrals \boldmath$D0_{b2}^{(1)}$ and
  $D0_{b2}^{(2)}$ \unboldmath}
\label{subsec:b2_scalar}

The scalar box integral $D0_{b2}^{(1)}$ can be written as:
\begin{equation} 
\label{eq:i_b2_integral}
D0_{b2}^{(1)}=\mu^{4-d}
\int\frac{d^dk}{(2\pi)^d}\frac{1}{N_1 N_2 N_3 N_4} \,\,\,,
\end{equation} 
where
\begin{eqnarray}
\label{eq:i_b2_denominators} 
N_1&=&k^2\,\,\,,\nonumber \\ 
N_2&=& (k+p_t)^2-m_t^2\,\,\,,\nonumber \\ 
N_3&=& (k+p_t+p_h)^2-m_t^2\,\,\,,\nonumber \\
N_4&=& (k-p_t^\prime)^2-m_t^2\,\,\,,
\end{eqnarray} 
while $D0_{b2}^{(2)}$ is obtained from $D0_{b2}^{(1)}$ by exchanging
$p_t\leftrightarrow p_t^\prime$. Therefore, all the following results for
$D0_{b2}^{(1)}$ can be easily extended to $D0_{b2}^{(2)}$.

The part of $D0_{b2}^{(1)}$ which contributes to the virtual amplitude
squared is of the form:
\begin{equation} 
\label{eq:i_b2_result}
D0_{b2}^{(1)}=\frac{i}{16\pi^2}{\cal N}_t\left( {X_{-1}\over
\epsilon}+X_0\right) {1\over (m_t^2-{\bar s}_{th}) {\bar s}_{t\bar t} 
\beta_{t\bar t}} \,\,\,,
\end{equation}
where ${\cal N}_t$ is given in Eq.~(\ref{eq:nsnt}), while
${\bar s}_{t\bar t}=(p_t+p_t^\prime)^2=(q-p_h)^2>0$,
$\beta_{t\bar t}=\sqrt{1-{4m_t^2\over {\bar s}_{t\bar t}}}$, 
${\bar s}_{th}=(p_t+p_h)^2$.
The pole part $X_{-1}$ is:
\begin{equation}
X_{-1}=\ln\left(\frac{1+\beta_{t\bar t}}{1-\beta_{t\bar t}}\right)\,\,\,,
\end{equation}
while the finite part can be calculated using Ref.~\cite{Beenakker:1990jr}.

\subsection{Box 3: box scalar integrals \boldmath$D0_{b3}^{(1)}$, 
$D0_{b3}^{(2)}$, $D0_{b3}^{(3)}$, and $D0_{b3}^{(4)}$\unboldmath} 
\label{subsec:b3_scalar}
The scalar box integral $D0_{b3}^{(1)}$ can be written as:
\begin{equation}
\label{eq:i_b3_integral}
D0_{b3}^{(1)}=\mu^{4-d}
\int\frac{d^dk}{(2\pi)^d} \frac{1}{N_1 N_2 N_3 N_4}\,\,\,,
\end{equation}
where
\begin{eqnarray}
\label{eq:i_b3_1_denominators}
N_1&=&k^2\,\,\,,\nonumber \\
N_2&=& (k+q_1)^2\,\,\,,\nonumber \\
N_3&=& (k+q_1+q_2)^2\,\,\,,\nonumber \\
N_4&=& (k+q_1+q_2-p_t^\prime)^2-m_t^2\,\,\, .
\end{eqnarray}
$D0_{b3}^{(2)}$ is obtained from $D0_{b3}^{(1)}$ by exchanging
$q_1\leftrightarrow q_2$. On the other hand,
$D0_{b3}^{(3)}$ arises from the box diagram where
the Higgs boson is emitted from the antitop quark and corresponds to:
\begin{eqnarray}
\label{eq:i_b3_3_denominators}
N_1&=&k^2\,\,\,,\nonumber \\
N_2&=& (k+q_2)^2\,\,\,,\nonumber \\
N_3&=& (k+q_1+q_2)^2\,\,\,,\nonumber \\
N_4&=& (k+q_1+q_2-p_t)^2-m_t^2\,\,\, .
\end{eqnarray}
Therefore $D0_{b3}^{(3)}$ can be obtained from $D0_{b3}^{(1)}$ by
exchanging $q_1\leftrightarrow q_2$ and $p_t^\prime\leftrightarrow
p_t$.  Finally, $D0_{b3}^{(4)}$ is obtained from $D0_{b3}^{(3)}$ by
exchanging $q_1\leftrightarrow q_2$.  We present here the case of
$D0_{b3}^{(1)}$. All other $D0_{b3}^{(i)}$ boxes, for $i\!=\!2,3,4$
can be obtained following the simple pattern of substitutions
explained above.

The part of $D0_{b3}^{(1)}$ which contributes to the virtual amplitude
squared is given by:
\begin{equation}
\label{eq:i_b3_result}
D0_{b3}^{(1)}=\frac{i}{16\pi^2}{\cal N}_t
\left(-\frac{1}{\sigma\tau_2}\right)
\left(\frac{X_{-2}}{\epsilon^2}+\frac{X_{-1}}{\epsilon}+X_0\right)\,\,\,,
\end{equation}
where ${\cal N}_t$ is defined in Eq.~(\ref{eq:nsnt}), the coefficients
$X_{-2}$, $X_{-1}$, and $X_{0}$ are given by:
\begin{eqnarray}
\label{eq:i_b3_coefficients}
X_{-2}&=&\frac{3}{2}\,\,\,, \\
X_{-1}&=&\ln\left(\frac{\omega_1 m_t^4}{\sigma \tau_2^2}\right)\,\,\,,
\nonumber \\
X_0&=&
2\ln\left(\frac{\tau_2}{m_t^2}\right)
\ln\left(\frac{\sigma}{m_t^2}\right)
-\ln^2\left(\frac{\omega_1}{m_t^2}\right)
 -2 \mbox{Li}_2\left(1+\frac{\omega_1}{\tau_2}\right)+\frac{\pi^2}{3}
\,\,\,,\nonumber 
\end{eqnarray}
and $\sigma$, $\tau_2$, and $\omega_1$ are defined in
Eq.~(\ref{eq:den_symbols}).

\section{Phase space soft integrals}
\label{sec:soft_int}

In this appendix we collect the integrals which we have used in
calculating the results in Eq.~(\ref{eq:sigma_soft_poles}) starting
from Eq.~(\ref{eq:sigma_soft}). For a more exhaustive treatment of the
formalism used we refer to Refs.~\cite{Harris:2001sx,Beenakker:1989bq}, from
which the results in this appendix have been taken.

We parameterize the soft gluon $d$-momentum in the $q\bar q$ rest
frame as:
\begin{equation}
\label{eq:gluon_param}
k=E_g(1,\ldots,\sin\theta_1\sin\theta_2, \sin\theta_1\cos\theta_2, 
\cos\theta_1)\,\,\,,
\end{equation} 
such that the phase space  of the soft gluon in $d\!=\!4-2\epsilon$
dimensions can be written as:
\begin{equation}
\label{eq:gluon_ps}
d(PS_g)_{soft}=\frac{\Gamma(1-\epsilon)}{\Gamma(1-2\epsilon)}
\frac{\pi^\epsilon}{(2\pi)^3} \int_0^{\delta_s \sqrt{s}/2} dE_g
E_g^{1-2\epsilon}
\int_0^{\pi} d \theta_1
\sin^{1-2\epsilon}\theta_1
\int_0^\pi d\theta_2 \sin^{-2\epsilon}\theta_2\,\,\,.
\end{equation}
Then, all the integrals we need are of the form:
\begin{equation}
\label{eq:in_general}
I_n^{(k,l)}=\int_0^\pi d\theta_1\sin^{d-3}\theta_1
\int_0^\pi d\theta_2\sin^{d-4}\theta_2
\frac{\left(a+b\cos\theta_1\right)^{-k}}
{\left(A+B\cos\theta_1+C\sin\theta_1\cos\theta_2\right)^l}\,\,\,.
\end{equation}
In particular we need the following four cases. When $A^2\neq
B^2+C^2$, and $b=-a$, we use (dropping terms of order ${\cal
O}\left((d-4)^2\right)$):
\begin{eqnarray}
I_n^{(1,1)}&=&\frac{\pi}{a(A+B)}
\left\{\frac{2}{d-4}+\ln\left[\frac{(A+B)^2}{A^2-B^2-C^2}\right]\right.\\
&+&\frac{1}{2}(d-4)\left[\ln^2\left(\frac{A-\sqrt{B^2+C^2}}{A+B}\right)
-\frac{1}{2}\ln^2\left(\frac{A+\sqrt{B^2+C^2}}{A-\sqrt{B^2+C^2}}\right)
\right.\nonumber\\
&+&\left.\left.
2\,\mbox{Li}_2\left(-\frac{B+\sqrt{B^2+C^2}}{A-\sqrt{B^2+C^2}}\right)
-2\,\mbox{Li}_2\left(\frac{B-\sqrt{B^2+C^2}}{A+B}\right)\right]\right\}\,\,\,,
\nonumber
\end{eqnarray}
while when $b\neq -a$ we use:
\begin{eqnarray}
\label{eq:in_01}
I_n^{(0,1)}&=&\frac{\pi}{\sqrt{B^2+C^2}}\left\{
\ln\left(\frac{A+\sqrt{B^2+C^2}}{A-\sqrt{B^2+C^2}}\right)\right.\\
&&\left.-(d-4)\left[
\mbox{Li}_2\left(\frac{2\sqrt{B^2+C^2}}{A+\sqrt{B^2+C^2}}\right)+
\frac{1}{4}\ln^2\left(\frac{A+\sqrt{B^2+C^2}}{A-\sqrt{B^2+C^2}}\right)
\right]\right\}\,\,\,,\nonumber
\end{eqnarray}
\begin{equation}
\label{eq:in_02}
I_n^{(0,2)}=\frac{2\pi}{A^2-B^2-C^2}
\left[1-\frac{1}{2}(d-4)\frac{A}{\sqrt{B^2+C^2}}
\ln\left(\frac{A+\sqrt{B^2+C^2}}{A-\sqrt{B^2+C^2}}\right)\right]\,\,\,.
\end{equation}
Finally, when $A^2=B^2+C^2$, and $b=-a$, we have:
\begin{equation}
\label{eq:in_11}
I_n^{(1,1)}=2\pi\frac{1}{aA}\,\frac{1}{d-4}\left(\frac{A+B}{2A}\right)^{d/2-3}
\left[1+\frac{1}{4}(d-4)^2\mbox{Li}_2\left(\frac{A-B}{2A}\right)\right]
\,\,\,.
\end{equation}

\section{Color ordered amplitudes for \boldmath 
$h\rightarrow q\bar qt\bar t+g$ \unboldmath}
\label{sec:a_real_a1a2a3a4}
The tree-level amplitude for $h\rightarrow q(q_1)\bar q(q_2)t(p_t)\bar
t(p_t^\prime)$ is explicitly given by:
\begin{eqnarray}
\label{eq:a_lo_a0}
{\cal A}^{h\to q\bar qt\bar t}_{\sss LO}&=&i\,\frac{m_t}{v}g_s^2
\delta_{f_qf_{\bar q}}
\delta_{f_tf_{\bar t}}\left[\bar u(q_1)\gamma^\nu T^a_{c_qc_{\bar q}}v(q_2)
\right]\frac{1}{(p_h-p_t-p_t^\prime)^2}\times\\
&&\left[\bar u(p_t)\left(\gamma_\nu
\frac{p_h\!\!\!\!\!/-p_t^\prime\!\!\!\!/+m_t}{(p_h-p_t^\prime)^2-m_t^2}+
\frac{-p_h\!\!\!\!\!/+p_t\!\!\!\!/+m_t}
{(p_h-p_t)^2-m_t^2}\gamma_\nu\right)
T^a_{c_tc_{\bar t}}v(p_t^\prime)\right]\nonumber\\
&=&\frac{1}{2}\left(
 \delta_{c_tc_{\bar q}}\delta_{c_qc_{\bar t}}-
\frac{1}{N}\delta_{c_tc_{\bar t}}\delta_{c_qc_{\bar q}}\right) 
\delta_{f_qf_{\bar q}}\delta_{f_tf_{\bar t}}{\cal A}_0\,\,\,,\nonumber
\end{eqnarray}
where $p_h$ is taken as incoming, while all the other momenta are
outgoing. Using the color decomposition given in Eq.~(\ref{eq:tata}),
we have rewritten ${\cal A}^{h\to q\bar qt\bar t}_{\sss LO}$ in terms
of a leading color and a sub-leading color ordered amplitude. Both
amplitudes are given by:
\begin{eqnarray}
\label{eq:a0}
{\cal A}_0&=&i\,\frac{m_t}{v}g_s^2 
\left[\bar u(q_1)\gamma^\nu v(q_2)\right]
\frac{1}{(p_h-p_t-p_t^\prime)^2}\times\\
&&\left[\bar u(p_t)\left(\gamma_\nu
\frac{p_h\!\!\!\!\!/-p_t^\prime\!\!\!\!/+m_t}{(p_h-p_t^\prime)^2-m_t^2}+
\frac{-p_h\!\!\!\!\!/+p_t\!\!\!\!/+m_t}
{(p_h-p_t)^2-m_t^2}\gamma_\nu\right)v(p_t^\prime)\right]
\nonumber\\
&=&i\,\frac{m_t}{v}g_s^2 {\cal A}^{0,\nu}_{q\bar q}
\frac{1}{(p_h-p_t-p_t^\prime)^2}
{\cal A}^0_{t\bar t,\nu}\,\,\,,\nonumber
\end{eqnarray}
where, for future purposes, we have introduced the ${\cal
A}^{0,\nu}_{q\bar q}$ and ${\cal A}^{0,\nu}_{t\bar t}$ tree-level
partial amplitudes:
\begin{eqnarray}
\label{eq:a0_partials}
{\cal A}^{0,\nu}_{q\bar q}&=&\bar u(q_1)\gamma^\nu v(q_2)\,\,\,,\\
{\cal A}^{0,\nu}_{t\bar t}&=&\bar u(p_t)\left(\gamma^\nu
\frac{p_h\!\!\!\!\!/-p_t^\prime\!\!\!\!/+m_t}{(p_h-p_t^\prime)^2-m_t^2}+
\frac{-p_h\!\!\!\!\!/+p_t\!\!\!\!/+m_t}{(p_h-p_t)^2-m_t^2}\gamma^\nu\right)
v(p_t^\prime)\,\,\,.\nonumber
\end{eqnarray}

The ${\cal O}(\alpha_s)$ real corrections to the Born amplitude
consist of the process $h\rightarrow q\bar qt\bar t+g$, where the
gluon can be emitted either from the external quark legs or from the
internal gluon propagator. Therefore we can write ${\cal A}^{h\to
q\bar qt\bar t g}$ as follows:
\begin{eqnarray}
\label{eq:a_real_sum}
{\cal A}^{h\to q\bar qt\bar t g}&=&(ig_s)\delta_{f_qf_{\bar q}}
\delta_{f_tf_{\bar t}}
\left[{\cal A}_q^\mu\,(T^aT^b)_{c_qc_{\bar q}}T^b_{c_tc_{\bar t}}
+{\cal A}_{\bar q}^\mu\,(T^bT^a)_{c_qc_{\bar q}}T^b_{c_tc_{\bar t}}\right.\\
&+& \left.
{\cal A}_t^\mu\,T^b_{c_qc_{\bar q}}\left(T^aT^b\right)_{c_tc_{\bar t}}+
{\cal A}_{\bar t}^\mu\,T^b_{c_qc_{\bar q}}\left(T^bT^a\right)_{c_tc_{\bar t}}+
{\cal A}_g^\mu\,(if^{abc}T^b_{c_tc_{\bar t}}T^c_{c_qc_{\bar q}})\right]
\cdot\epsilon_\mu(k)\,\,\,,\nonumber
\end{eqnarray}
where $\epsilon^\mu(k)$ is the polarization vector of the emitted
gluon and we have defined by ${\cal A}_i^\mu$ the part of the real
amplitude corresponding to the emission of the gluon from
$i\!=\!q,\bar q,t,\bar t,g$. More explicitly, the ${\cal A}_i^\mu$
amplitudes are given by:
\begin{eqnarray}
\label{eq:a_real_qq_tt_g}
{\cal A}_{q}^\mu&=&\left(g_s^2\frac{m_t}{v}\right)\left(\bar u(q_1)
\gamma^\mu\frac{q_1\!\!\!\!\!/+k\!\!\!/}{2q_1\cdot k}\gamma_\nu v(q_2)\right)
\frac{1}{(p_h-p_t-p_t^\prime)^2} {\cal A}_{t\bar t}^{0,\nu}\,\,\,,\\
{\cal A}_{\bar q}^\mu&=&\left(g_s^2\frac{m_t}{v}\right)\left(\bar u(q_1)
\gamma_\nu\frac{-q_2\!\!\!\!\!/-k\!\!\!/}{2q_2\cdot k}\gamma^\mu
v(q_2)\right)\frac{1}{(p_h-p_t-p_t^\prime)^2}
{\cal A}_{t\bar t}^{0,\nu}\,\,\,,\nonumber\\
{\cal A}_g^\mu&=&\left(g_s^2\frac{m_t}{v}\right){\cal A}^{0}_{q\bar q,\rho}
\frac{1}{(p_h-p_t-p_t^\prime)^2}\left(V_{3g}^{\mu\rho\nu}(k,q_1,q_2)
\right)
\frac{1}{(q_1+q_2)^2} {\cal A}^{0}_{t\bar t,\nu}\,\,\,,\nonumber\\
{\cal A}_{t}^\mu&=&\left(g_s^2\frac{m_t}{v}\right)
{\cal A}^{0,\nu}_{q\bar q}\bar u(p_t)\left(
\gamma^\mu\frac{p_t\!\!\!\!/+k\!\!\!/+m_t}{2p_t\cdot k}\gamma_\nu
\frac{p_h\!\!\!\!\!/-p_t^\prime\!\!\!\!/+m_t}{(p_h-p_t^\prime)^2-m_t^2}
\right.
\nonumber\\
&+&
\frac{-p_h\!\!\!\!\!/+p_t\!\!\!\!/+m_t}{(p_h-p_t)^2-m_t^2}\gamma^\mu
\frac{-p_h\!\!\!\!\!/+p_t\!\!\!\!/+k\!\!\!/+m_t}{(p_h-p_t-k)^2-m_t^2}
\gamma_\nu\nonumber\\
&+&\left.\gamma^\mu
\frac{p_t\!\!\!\!/+k\!\!\!/+m_t}{2p_t\cdot k}\,
\frac{-p_h\!\!\!\!\!/+p_t\!\!\!\!/+k\!\!\!/+m_t}{(p_h-p_t-k)^2-m_t^2}
\gamma_\nu
\right)\frac{1}{(q_1+q_2)^2}v(p_t^\prime)\,\,\,,\nonumber\\
{\cal A}_{\bar t}^\mu&=&\left(g_s^2\frac{m_t}{v}\right)
{\cal A}^{0,\nu}_{q\bar q}\bar u(p_t)
\left(
\frac{-p_h\!\!\!\!\!/+p_t\!\!\!\!/+m_t}{(p_h-p_t)^2-m_t^2}\gamma_\nu
\frac{-p_t^\prime\!\!\!\!/-k\!\!\!/+m_t}{2p_t^\prime\cdot k}\gamma^\mu
\right.\nonumber\\
&+&
\gamma_\nu\frac{p_h\!\!\!\!\!/-p_t^\prime\!\!\!\!/-k\!\!\!/+m_t}
{(p_h-p_t^\prime-k)^2-m_t^2}\gamma^\mu
\frac{p_h\!\!\!\!\!/-p_t^\prime\!\!\!\!/+m_t}{(p_h-p_t^\prime)^2-m_t^2}
\nonumber\\
&+&\left.
\gamma_\nu\frac{p_h\!\!\!\!\!/-p_t\!\!\!\!/-k\!\!\!/+m_t}
{(p_h-p_t-k)^2-m_t^2}\,
\frac{-p_t^\prime\!\!\!\!/-k\!\!\!/+m_t}{2p_t^\prime\cdot k}\gamma^\mu
\right)
\frac{1}{(q_1+q_2)^2}v(p_t^\prime)\,\,\,,\nonumber
\end{eqnarray}
where
\begin{equation}
\label{eq:v3g}
V_{3g}^{\mu\rho\nu}(k,q_1,q_2)=(-2k^\rho-q^\rho)g^{\mu\nu}+
(2q^\mu+k^\mu)g^{\nu\rho}+(-q^\nu-k^\nu)g^{\mu\rho})\,\,\,.
\end{equation}
Using the color decomposition given in Eq.~(\ref{eq:tata}), we can
also rewrite ${\cal A}^{h\rightarrow q\bar qt\bar tg}$ as a
linear combination of four color ordered amplitudes, as already given
in Eq.~(\ref{eq:a_real_a1a2a3a4}). By matching the color factors in
Eq.~(\ref{eq:a_real_sum}) to the color factors in
Eq.~(\ref{eq:a_real_a1a2a3a4}), we see that the color ordered
amplitudes ${\cal A}_i(q_1,q_2,p_t,p_t^\prime,k)$ (for
$i\!=\!1,\dots,4$) are given by \cite{Berends:1989yn}:
\begin{eqnarray}
\label{eq:a1a2a3a4}
{\cal A}_1(q_1,q_2,p_t,p_t^\prime,k)&=&\left({\cal A}_{q}^\mu+
{\cal A}_{\bar t}^\mu-{\cal A}_g^\mu\right)\cdot\epsilon_\mu(k)\,\,\,,\\
{\cal A}_2(q_1,q_2,p_t,p_t^\prime,k)&=&\left({\cal A}_{\bar q}^\mu+
{\cal A}_{t}^\mu+{\cal A}_g^\mu\right)\cdot\epsilon_\mu(k)\,\,\,,\nonumber\\
{\cal A}_3(q_1,q_2,p_t,p_t^\prime,k)&=&\left({\cal A}_{q}^\mu+
{\cal A}_{\bar q}^\mu\right)\cdot\epsilon_\mu(k)\,\,\,,\nonumber\\
{\cal A}_4(q_1,q_2,p_t,p_t^\prime,k)&=&\left({\cal A}_{t}^\mu+
{\cal A}_{\bar t}^\mu\right)\cdot\epsilon_\mu(k)\,\,\,.\nonumber
\end{eqnarray}
\end{document}